\documentclass[aps,prc,twocolumn,superscriptaddress,showpacs,nofootinbib,floatfix]{revtex4}
\usepackage{graphicx}
\usepackage{dcolumn}
\usepackage{bm}
\newcommand{\pp} {\mbox{$p+p$}}
\newcommand{\nn} {\mbox{$N+N$}}
\newcommand{\Au} {\mbox{$Au+Au$}}
\newcommand{\pt} {\mbox{$p_T$}}
\newcommand{\xt} {\mbox{$x_T$}}

\newcommand{\raa} {\mbox{$R_{AA}$}}
\newcommand{\raap} {\mbox{$R_{AA}^{N_{part}}$}}
\newcommand{\snn} {\mbox{$\sqrt{s_{NN}}$}}

\newcommand{\npmt} {\mbox{$N_{PMT}$}}
\newcommand{\meanpmt} {\mbox{$\langle N_{PMT} \rangle$}}
\newcommand{\taa} {\mbox{$\langle T_{AuAu} \rangle$}}
\newcommand{\trun}  {\mbox{$\langle p_T^{trunc} \rangle$}}
\newcommand{\piz} {\mbox{$\pi^{0}$}}

\newcommand{\npart} {\mbox{$N_{\it part}$}}
\newcommand{\ncoll} {\mbox{$N_{\it coll}$}}
\newcommand{\anpart} {\mbox{$\langle N_{\it part} \rangle$}}
\newcommand{\ancoll} {\mbox{$\langle N_{\it coll} \rangle$}}
\newcommand{\brf} {\mbox{$D_{\phi}^+$}}

\newcommand{\naf} {\mbox{$D_{\phi}^-$}}

\def\Journal#1#2#3#4{{#1}{\bf #2}, #3 (#4)}

\def\EPJ{{Eur. Phys. J.}~{\bf C}}
\def\JHEP{{J. High Energy Phys.}~}

\def\NIMA{{Nucl. Instrum. Methods}~{\bf A}}
\def\NPA{{Nucl. Phys.}~{\bf A}}
\def\NPB{{Nucl. Phys.}~{\bf B}}
\def\PLB{{Phys. Lett.}~{\bf B}}

\def\PRL{Phys. Rev. Lett.\ }
\def\PRD{{Phys. Rev.}~{\bf D}}
\def\PRC{{Phys. Rev.}~{\bf C}}
\def\ZPC{{Z. Phys.}~{\bf C}}
\def\ARNS{{Ann. Rev. Nucl. Part. Sci.\ }}
\def\RMP{Rev. Mod. Phys.\ }
\def\LNP{Lect. Notes. Phys.\ }

\begin{document}
\title{High-$\pt$ Charged Hadron Suppression in $\Au$ Collisions at $\snn$ = 200 GeV}

\newcommand{\abilene}{Abilene Christian University, Abilene, TX 79699, USA}
\newcommand{\acadsin}{Institute of Physics, Academia Sinica, Taipei 11529, Taiwan}
\newcommand{\banaras}{Department of Physics, Banaras Hindu University, Varanasi 221005, India}
\newcommand{\barc}{Bhabha Atomic Research Centre, Bombay 400 085, India}
\newcommand{\bnl}{Brookhaven National Laboratory, Upton, NY 11973-5000, USA}
\newcommand{\caucr}{University of California - Riverside, Riverside, CA 92521, USA}
\newcommand{\ciae}{China Institute of Atomic Energy (CIAE), Beijing, People's Republic of China}
\newcommand{\cns}{Center for Nuclear Study, Graduate School of Science, University of Tokyo, 7-3-1 Hongo, Bunkyo, Tokyo 113-0033, Japan}
\newcommand{\columbia}{Columbia University, New York, NY 10027 and Nevis Laboratories, Irvington, NY 10533, USA}
\newcommand{\dapnia}{Dapnia, CEA Saclay, F-91191, Gif-sur-Yvette, France}
\newcommand{\debrecen}{Debrecen University, H-4010 Debrecen, Egyetem t{\'e}r 1, Hungary}
\newcommand{\fsu}{Florida State University, Tallahassee, FL 32306, USA}
\newcommand{\gsu}{Georgia State University, Atlanta, GA 30303, USA}
\newcommand{\hiroshima}{Hiroshima University, Kagamiyama, Higashi-Hiroshima 739-8526, Japan}
\newcommand{\ihepprot}{Institute for High Energy Physics (IHEP), Protvino, Russia}
\newcommand{\isu}{Iowa State University, Ames, IA 50011, USA}
\newcommand{\jinrdubna}{Joint Institute for Nuclear Research, 141980 Dubna, Moscow Region, Russia}
\newcommand{\kaeri}{KAERI, Cyclotron Application Laboratory, Seoul, South Korea}
\newcommand{\kangnung}{Kangnung National University, Kangnung 210-702, South Korea}
\newcommand{\kek}{KEK, High Energy Accelerator Research Organization, Tsukuba-shi, Ibaraki-ken 305-0801, Japan}
\newcommand{\kfki}{KFKI Research Institute for Particle and Nuclear Physics (RMKI), H-1525 Budapest 114, POBox 49, Hungary}
\newcommand{\korea}{Korea University, Seoul, 136-701, Korea}
\newcommand{\kurchatov}{Russian Research Center ``Kurchatov Institute", Moscow, Russia}
\newcommand{\kyoto}{Kyoto University, Kyoto 606, Japan}
\newcommand{\labllr}{Laboratoire Leprince-Ringuet, Ecole Polytechnique, CNRS-IN2P3, Route de Saclay, F-91128, Palaiseau, France}
\newcommand{\lawllnl}{Lawrence Livermore National Laboratory, Livermore, CA 94550, USA}
\newcommand{\losalamos}{Los Alamos National Laboratory, Los Alamos, NM 87545, USA}
\newcommand{\lpc}{LPC, Universit{\'e} Blaise Pascal, CNRS-IN2P3, Clermont-Fd, 63177 Aubiere Cedex, France}
\newcommand{\lund}{Department of Physics, Lund University, Box 118, SE-221 00 Lund, Sweden}
\newcommand{\muenster}{Institut fuer Kernphysik, University of Muenster, D-48149 Muenster, Germany}
\newcommand{\myongji}{Myongji University, Yongin, Kyonggido 449-728, Korea}
\newcommand{\nagasaki}{Nagasaki Institute of Applied Science, Nagasaki-shi, Nagasaki 851-0193, Japan}
\newcommand{\newmex}{University of New Mexico, Albuquerque, NM 87131, USA}
\newcommand{\nmsu}{New Mexico State University, Las Cruces, NM 88003, USA}
\newcommand{\ornl}{Oak Ridge National Laboratory, Oak Ridge, TN 37831, USA}
\newcommand{\orsay}{IPN-Orsay, Universite Paris Sud, CNRS-IN2P3, BP1, F-91406, Orsay, France}
\newcommand{\pnpi}{PNPI, Petersburg Nuclear Physics Institute, Gatchina, Russia}
\newcommand{\riken}{RIKEN (The Institute of Physical and Chemical Research), Wako, Saitama 351-0198, JAPAN}
\newcommand{\rkrbrc}{RIKEN BNL Research Center, Brookhaven National Laboratory, Upton, NY 11973-5000, USA}
\newcommand{\saispbstu}{St. Petersburg State Technical University, St. Petersburg, Russia}
\newcommand{\saopaulo}{Universidade de S{\~a}o Paulo, Instituto de F\'{\i}sica, Caixa Postal 66318, S{\~a}o Paulo CEP05315-970, Brazil}
\newcommand{\seoulnat}{System Electronics Laboratory, Seoul National University, Seoul, South Korea}
\newcommand{\stonybrkc}{Chemistry Department, Stony Brook University, SUNY, Stony Brook, NY 11794-3400, USA}
\newcommand{\stonycrkp}{Department of Physics and Astronomy, Stony Brook University, SUNY, Stony Brook, NY 11794, USA}
\newcommand{\subatech}{SUBATECH (Ecole des Mines de Nantes, CNRS-IN2P3, Universit{\'e} de Nantes) BP 20722 - 44307, Nantes, France}
\newcommand{\tenn}{University of Tennessee, Knoxville, TN 37996, USA}
\newcommand{\titech}{Department of Physics, Tokyo Institute of Technology, Tokyo, 152-8551, Japan}
\newcommand{\tsukuba}{Institute of Physics, University of Tsukuba, Tsukuba, Ibaraki 305, Japan}
\newcommand{\vandy}{Vanderbilt University, Nashville, TN 37235, USA}
\newcommand{\waseda}{Waseda University, Advanced Research Institute for Science and Engineering, 17 Kikui-cho, Shinjuku-ku, Tokyo 162-0044, Japan}
\newcommand{\weizmann}{Weizmann Institute, Rehovot 76100, Israel}
\newcommand{\yonsei}{Yonsei University, IPAP, Seoul 120-749, Korea}
\affiliation{\abilene}
\affiliation{\acadsin}
\affiliation{\banaras}
\affiliation{\barc}
\affiliation{\bnl}
\affiliation{\caucr}
\affiliation{\ciae}
\affiliation{\cns}
\affiliation{\columbia}
\affiliation{\dapnia}
\affiliation{\debrecen}
\affiliation{\fsu}
\affiliation{\gsu}
\affiliation{\hiroshima}
\affiliation{\ihepprot}
\affiliation{\isu}
\affiliation{\jinrdubna}
\affiliation{\kaeri}
\affiliation{\kangnung}
\affiliation{\kek}
\affiliation{\kfki}
\affiliation{\korea}
\affiliation{\kurchatov}
\affiliation{\kyoto}
\affiliation{\labllr}
\affiliation{\lawllnl}
\affiliation{\losalamos}
\affiliation{\lpc}
\affiliation{\lund}
\affiliation{\muenster}
\affiliation{\myongji}
\affiliation{\nagasaki}
\affiliation{\newmex}
\affiliation{\nmsu}
\affiliation{\ornl}
\affiliation{\orsay}
\affiliation{\pnpi}
\affiliation{\riken}
\affiliation{\rkrbrc}
\affiliation{\saispbstu}
\affiliation{\saopaulo}
\affiliation{\seoulnat}
\affiliation{\stonybrkc}
\affiliation{\stonycrkp}
\affiliation{\subatech}
\affiliation{\tenn}
\affiliation{\titech}
\affiliation{\tsukuba}
\affiliation{\vandy}
\affiliation{\waseda}
\affiliation{\weizmann}
\affiliation{\yonsei}
\author{S.S.~Adler} \affiliation{\bnl}
\author{S.~Afanasiev}   \affiliation{\jinrdubna}
\author{C.~Aidala}  \affiliation{\bnl}
\author{N.N.~Ajitanand} \affiliation{\stonybrkc}
\author{Y.~Akiba}   \affiliation{\kek} \affiliation{\riken}
\author{J.~Alexander}   \affiliation{\stonybrkc}
\author{R.~Amirikas}    \affiliation{\fsu}
\author{L.~Aphecetche}  \affiliation{\subatech}
\author{S.H.~Aronson}   \affiliation{\bnl}
\author{R.~Averbeck}    \affiliation{\stonycrkp}
% ...
\collaboration{PHENIX Collaboration} \noaffiliation

\date{\today}
\begin{abstract}
The PHENIX experiment at RHIC has measured charged hadron yields
at mid-rapidity over a wide range of transverse momentum ($0.5
<\pt <10$~GeV/$c$) in $\Au$ collisions at $\snn$ = 200 GeV. The
data are compared to $\piz$ measurements from the same experiment.
For both charged hadrons and neutral pions, the yields per
nucleon-nucleon collision are significantly suppressed in central
compared to peripheral and nucleon-nucleon collisions. The
suppression sets in gradually and increases with increasing
centrality of the collisions. Above 4-5 GeV/$c$ in $\pt$, a
constant and almost identical suppression of charged hadrons and
$\piz$'s is observed.
The $\pt$ spectra are compared to published
spectra from $\Au$ at $\snn$ = 130 in terms of $\xt$ scaling.
Central and peripheral $\piz$ as well as peripheral charged
spectra exhibit the same $\xt$ scaling as observed in $\pp$ data.
\end{abstract}
\maketitle


\section{INTRODUCTION}
\label{sec:intro}

Lattice Quantum-Chromo-Dynamic (QCD) calculations predict  a new
state of matter of deconfined quarks and gluons at an energy
density exceeding $\sim$1~GeV/$fm^3$~\cite{latt01}.  It has long
been suggested that such a ``quark gluon plasma'' may be produced
in collisions between ultra-relativistic heavy
nuclei~\cite{latt02}. Indeed, measurements of transverse energy
produced in high energy $Pb+Pb$ and $\Au$ collision suggest that
energy densities above 3~GeV/$fm^3$ at SPS~\cite{na49} and
5~GeV/$fm^3$ at RHIC~\cite{ppg002,sasha} have been reached.
However, this conclusion relies on model
assumptions~\cite{bjorken,Teaney,Kolb,Florkowski} to relate the
properties of the hadronic final state to the initial state
dynamics.

The spectra of high transverse momentum ($\pt$) hadrons resulting
from the fragmentation of hard-scattered partons potentially
provide a direct probe of the properties of the initial state.
Theoretical calculations show that the outgoing high-$\pt$ partons
radiate substantially more energy when propagating through dense
matter than when propagating in the vacuum, resulting in a
softening of the hadron $\pt$ spectrum~\cite{Gyu90}, with the
energy loss of the partons depending on the gluon density of the
matter~\cite{BDMPS,baier}. Formation time considerations suggest
that hard scattered partons are ``produced'' at the earliest stage
of the collision, thus directly probe the dense matter from the
time of their creation. Therefore, a detailed analysis of
high-$\pt$ hadron production may reveal information on the
properties of the dense medium created early in the collisions
~\cite{xnwang1,xnwang2,baier}.

At the energies reached at the Relativistic Heavy Ion Collider
(RHIC), high-$\pt$ hadrons are copiously produced. In
nucleon-nucleon collisions, it has been well established that
hadrons with $\pt\geq$ 2 GeV/$c$ result primarily from the
fragmentation of hard-scattered partons, and that the $\pt$
spectra of these hadrons can be calculated using perturbative QCD
(pQCD)~\cite{owens1,ppg024}. Initial measurements of hadron $\pt$
spectra in $\Au$ collisions at $\snn$ = 130 GeV led to the
discovery of a substantial suppression of hadron yields per
nucleon-nucleon collision relative to pp
data~\cite{ppg003,star,ppg013}. Data from $\snn$ = 200 GeV confirm
these results~\cite{phobosch,ppg014,starch,brahmsch}. The
suppression is observed in central but not in peripheral
collisions. These observations are consistent with pQCD-inspired
modelling of parton energy loss in dense
matter~\cite{xnwang3,levai}. However, alternative interpretations
that do not assume the formation of a deconfined phase have been
proposed based on the modifications of the parton distribution
functions in the initial state~\cite{dima} or final-state hadronic
interactions~\cite{gallmeister}.

In addition to hadron suppression, an unexpectedly large fraction
of baryons has been observed in central $\Au$ collisions for $\pt$
up to 4--5~GeV/$c$~\cite{ppg006,ppg015,starks}, which complicates
the interpretation of the high $\pt$ results. The observed baryon
to meson ratio from PHENIX~\cite{ppg015} is inconsistent with jet
fragmentation in $\pp$~\cite{ISR} and $e^+e^-$
collisions~\cite{delphi}. While the origin of this effect is
unclear, it could point towards bulk particle production (``soft
physics'') contributing to the $\pt$ spectra out to 4--5 GeV/$c$.
It has been suggested that coalescence of thermalized quarks
combining with energy loss of hard-scattered partons can account
for the unusual particle composition, which shifts the region
dominated by hard-scattering to higher $\pt$~\cite{coalence}.

Systematic measurements of the $\pt$, centrality, particle
species, and $\snn$ dependence of the suppression can constrain
competing descriptions of high-$\pt$ hadron production. In this
paper, we present new data on inclusive charged hadron production
for $0.5 < \pt <10$~GeV/$c$, measured over a broad range of
centrality in $\Au$ collisions at $\snn$ = 200~GeV by the PHENIX
Collaboration at RHIC. These data are compared to data on neutral
pion production \cite{ppg014} and to data from $\Au$ collisions at
$\snn$ = 130~GeV \cite{ppg003,ppg013}, all measured within the
same experiment.

The remainder of the paper is organized as follows. Section
~\ref{sec:Sexpandana} gives a detailed account of the charged
particle analysis. Centrality and $\pt$ dependence of the charged
hadron $\pt$ spectra are discussed in Section ~\ref{sec:Scharge}.
Section ~\ref{sec:Scp_pi0} studies the charged hadron suppression
and compares the results to $\piz$ data. In Section
~\ref{sec:SxT}, we discuss the $\snn$ dependence of both charged
hadron and neutral pion production and test possible
$\xt$-scaling. A summary is given in Section~\ref{sec:conclusion}.

\section{DATA ANALYSIS}
\label{sec:Sexpandana}
\subsection{PHENIX Detector}
\label{sec:Sexp}

\begin{figure}[t]
\includegraphics[width=1.0\linewidth]{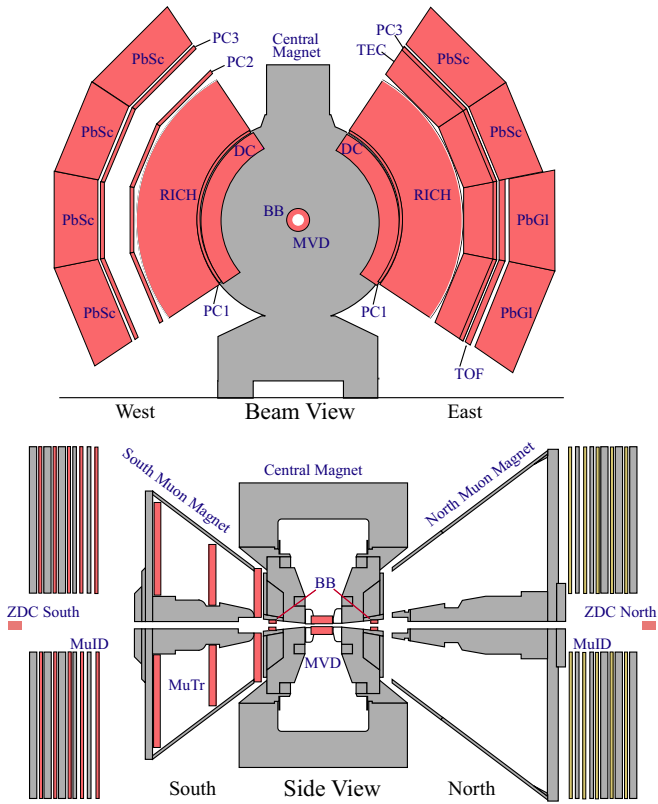}
     \caption{\label{fig:phenix}
    (Color online) PHENIX experimental layout for the $\Au$ run in 2001. The top panel
shows the PHENIX central arm spectrometers viewed along the beam
axis. The bottom panel shows a side view of the PHENIX muon arm
spectrometers.
    }
\end{figure}

The PHENIX experiment consists of four spectrometer arms $-$ two
around mid-rapidity (the central arms) and two at forward rapidity
(the muon arms) $-$ and a set of global detectors. The central arm
and south Muon arm detectors were completed in 2001 and took data
during $\Au$ operation of RHIC the same year (RUN-2). The layout
of the PHENIX experiment during RUN-2 is shown in
Figure~\ref{fig:phenix}. Each central arm covers the
pseudo-rapidity range $|\eta| < 0.35$ and 90 degrees in azimuthal
angle $\phi$. In each of the central arms, charged particles are
tracked by a drift chamber (DC) positioned from 2.0 to 2.4m
radially outward from the beam axis and 2 or 3 layers of pixel pad
chambers (PC1, (PC2), PC3 located at 2.4m, (4.2m), 5m in radial
direction, respectively). Particle identification is provided by
ring imaging Cerenkov counters (RICH), a time of flight
scintillator wall (TOF), and two types of electromagnetic
calorimeters (lead scintillator (PBSC) and lead glass (PBGL)). The
magnetic field for the central spectrometer is axially symmetric
around the beam axis. Its component parallel to the beam axis has
an approximately Gaussian dependence on the radial distance from
the beam axis, dropping from 0.48~T at the center to 0.096~T
(0.048~T) at the inner (outer) radius of the DC. A pair of
Zero-Degree Calorimeters (ZDC) and a pair of Beam-Beam Counters
(BBC) were used for global event characterization. Further details
about the design and performance of PHENIX can be found
in~\cite{phenixnim}.

\subsection{Event Selection}
\label{sec:evts}

During RUN-2, PHENIX sampled an integrated luminosity of $24~\mu
b^{-1}$ for $\Au$ collisions at $\snn$ = 200 GeV.  Minimum bias
events were selected by a coincidence between the ZDCs and the
BBCs. This selection corresponds to $92.2^{+2.5}_{-3.0}\%$ of the
6.9 barn $\Au$ inelastic cross section. The event centrality is
determined by correlating the charge detected in the BBCs with the
energy measured in the ZDCs. Two sets of centrality definitions
are used in this analysis: a ``{\it Fine}'' set of centralities,
which corresponds to 0-5\%, ...,15-20\%,20-30\%,...,80-92\%, and a
``{\it Coarse}'' set of centralities, which corresponds to
0-10\%,10-20\%,20-30\%,...,80-92\%. A Glauber model Monte-Carlo
simulation~\cite{glauber,foot0,ppg001,ppg009} that includes the
responses of BBC and ZDC gives an estimate of the average number
of binary collisions $\ancoll$, participating nucleons $\anpart$
and nuclear overlap function $\taa$ for each centrality class. The
calculated values of $\ancoll$, $\anpart$ and $\taa$ for each
centrality class are listed in Table~\ref{tab:sys0}.

In addition to the event selection, the BBCs also allow us to
reconstruct the collision vertex in the beam direction ($z$) with
a resolution of 0.5~$cm$. An offline $z$-vertex cut, $|z_{vtx}| <
$30~$cm$, was applied to the minimum bias events. After this
selection, a total of $27 \times 10^6$ minimum bias $\Au$ events
were analyzed to obtain the charged hadron spectra presented in
this paper.

\begin{table}
\caption{\label{tab:sys0} Centrality classes, average number of
$\nn$ collisions, average number of participant nucleons, and
average nuclear overlap function obtained from a Glauber
Monte-Carlo simulation of the BBC and ZDC responses for $\Au$ at
$\snn$ = 200 GeV. Each centrality class is expressed as a
percentage of $\sigma_{AuAu}$ = 6.9 b. Two sets of centrality
definitions are used in this analysis: a ``{\it Fine}'' set of
centralities, which corresponds to 0-5\%,
...,15-20\%,20-30\%,...,80-92\%, and a ``{\it Coarse}'' set of
centralities, which corresponds to
0-10\%,10-20\%,20-30\%,...,80-92\%.}
\begin{ruledtabular} \begin{tabular}{cccc}
Centrality     &       $\ancoll$                  &$\anpart$ &  $\taa (mb^{-1})$   \\ \hline
 0 - 5\%       &       $1065  \pm 105.5$       &       $351.4 \pm 2.9$&  $25.37 \pm 1.77$      \\
 5 - 10\%      &       $854.4 \pm 82.1$        &       $ 299  \pm 3.8$&  $20.13 \pm 1.36$      \\
10 - 15\%      &       $672.4 \pm 66.8$        &       $253.9 \pm 4.3$&  $16.01 \pm 1.15$      \\
15 - 20\%      &       $532.7 \pm 52.1$        &       $215.3 \pm 5.3$&  $12.68 \pm 0.86$      \\\hline
 0 - 10\%      &       $955.4 \pm 93.6$        &       $325.2 \pm 3.3$&  $22.75 \pm 1.56$      \\
10 - 20\%      &       $602.6 \pm 59.3$        &       $234.6 \pm 4.7$&  $14.35 \pm 1.00$      \\
20 - 30\%      &       $373.8 \pm 39.6$        &       $166.6 \pm 5.4$&  $ 8.90 \pm 0.72$      \\
30 - 40\%      &       $219.8 \pm 22.6$        &       $114.2 \pm 4.4$&  $ 5.23 \pm 0.44$      \\
40 - 50\%      &       $120.3 \pm 13.7$        &       $ 74.4 \pm 3.8$&  $ 2.86 \pm 0.28$      \\
50 - 60\%      &       $ 61.0 \pm 9.9$         &       $ 45.5 \pm 3.3$&  $ 1.45 \pm 0.23$      \\
60 - 70\%      &       $ 28.5 \pm 7.6$         &       $ 25.7 \pm 3.8$&  $ 0.68 \pm 0.18$      \\
70 - 80\%      &       $ 12.4 \pm 4.2$         &       $ 13.4 \pm 3.0$&  $ 0.30 \pm 0.10$      \\
80 - 92\%      &       $  4.9 \pm 1.2$         &       $  6.3 \pm 1.2$&  $ 0.12 \pm 0.03$      \\\hline
60 - 92\%      &       $ 14.5 \pm 4$           &       $ 14.5 \pm 2.5$&  $ 0.35 \pm 0.10$      \\\hline
min. bias      &       $257.8 \pm 25.4$        &       $109.1 \pm 4.1$&  $ 6.14 \pm 0.45$      \\
\end{tabular}  \end{ruledtabular}
\end{table}

\subsection{Charged Particle Tracking and Momentum Measurement}
\label{sec:Sreco}

Charged hadron tracks are measured using information from the DC,
PC1, PC2 and PC3 detectors of the west central-arm and the BBC.
The projections of the charged particle trajectories into a plane
perpendicular to the beam axis are detected typically in 12 wire
planes in the DC. The wire planes spaced at 0.6~$cm$ intervals
along the radial direction from the beam axis. Each wire provides
a projective measurement, with better than 150 $\mu m$ spacial
resolution in the azimuthal ($\phi$) direction. Eight additional
wire planes in the DC provide stereoscopic projections, which
together with the space point measured at the PC1 and the vertex
position measured by the BBC determine the polar angle of the
track. Trajectories are confirmed by requiring matching hits at
both PC2 and PC3 to reduce the secondary background.

Tracks are then projected back to the collision vertex through the
magnetic field to determine the momentum $\vec{p}$. The transverse
momentum $\pt$ is related to the deflection angle $\alpha$
measured at the DC with respect to an infinite momentum
trajectory. For tracks emitted perpendicular to the beam axis,
this relation can be approximated by
\begin{equation}
    \alpha \simeq \frac{K}{\pt}\;\qquad , \label{eq:1}
\end{equation}
where $K$ = 87 mrad GeV/$c$ is the effective field integral.

The momentum scale is verified by comparing the known proton mass
to the value measured for charged particles identified as protons
from their time-of-flight. The flight-time is measured in the TOF
detector, which cover $\pi/4$ of the azimuthal acceptance in the
east arm. The absolute value of the momentum scale is known to be
correct to better than 0.7\%.

The momentum resolution is directly related to the $\alpha$
resolution,
\begin{eqnarray}
    \delta p/p  &=& \delta \alpha/\alpha  \\
        &=&\frac{1}{K}\sqrt{(\frac{\sigma_{ms}}{\beta})^2
                  + (\sigma_{\alpha}p)^2} \qquad,\nonumber
\label{eq:reso}
\end{eqnarray}
where $\delta\alpha$ is the measured angular spread, which can be
decomposed into the contribution from multiple scattering
($\sigma_{ms}$) and the contribution from the intrinsic pointing
resolution ($\sigma_{\alpha}$) of the DC. At high $\pt$,
$\sigma_{\alpha}$ is the dominating contribution, i.e.
$\delta_{\alpha} \simeq \sigma_{\alpha}$. We measure
$\sigma_{\alpha} \approx 0.84 \pm 0.05$ mrad/(GeV/$c$) using zero
field data, where we select high-momentum tracks by requiring
energetic hadronic showers in the electromagnetic calorimeters.
The width of the proton mass as function of $\pt$ independently
confirms the momentum resolution. In summary, the momentum
resolution is determined to be $\delta p/p \simeq 0.7\% \oplus
1.0\%\ p$~(GeV/$c$). Further details on track reconstruction and
momentum determination can be found in~\cite{tracknim}.
\subsection{Background Rejection and Subtraction}
\label{sec:Sbg}

Approximately 95\% of the tracks reconstructed by the DC originate
from the event vertex. The remainder have to be investigated as
potential background to the charged particle measurement. The main
background sources include secondary particles from decays and
$e^{+}e^{-}$ pairs from the conversion of photons in materials
between the vertex and the DC. Depending on how close the
conversion or decay point is to the DC, or depending on the
Q-value of the decay, these tracks may have a small deflection
angle $\alpha$ at the DC. Thus, according to Eq.~\ref{eq:1}, they
are incorrectly assigned a large momentum. In this analysis, the
$\pt$ range over which charged particle production is accessible
in PHENIX is limited by this background. We exploit the track
match to PC2 and PC3 to reject as much of the background as
possible, then employ a statistical method to measure and subtract
the irreducible background.

For primary tracks, the distance in both the r-$\phi$ and the $z$
direction between the track projection point and the measured PC
hit position is approximately Gaussian with a mean of 0 and a
width given by,
\begin{equation}
    \sigma_{match} =
    \sqrt{{\sigma^{match}_{det}}^{2}+\left(\frac{\sigma^{match}_{ms}}{p\beta}\right)^{2}} \qquad ,
\label{eq:2}
\end{equation}
where $\sigma^{match}_{det}$ is the finite detector resolution
(which includes DC pointing (or $\alpha$) resolution and the PC2,
PC3 spacial resolution), and $\sigma^{match}_{ms}$ is the multiple
scattering contribution.

\begin{figure}[t]
\includegraphics[width=1.0\linewidth]{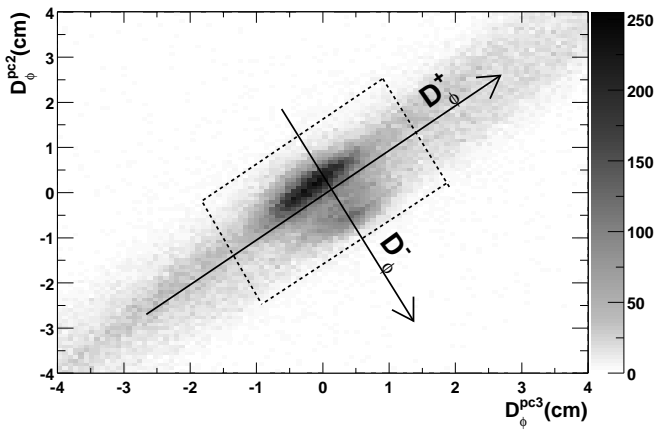}
     \caption{\label{fig:pc2pc3}
    $D_{\phi}^{pc2}$ (the difference between projection and hit location in
r-$\phi$ direction at PC2) versus $D_{\phi}^{pc3}$ in centimeters
for tracks with reconstructed $\pt > $ 4 GeV/$c$. PC2, PC3
matching differences are correlated, with signal tracks peaked
around 0 and background tracks extend along the $D_{\phi}^+$
direction. The double-peak structure along $D_{\phi}^-$ is related
to the finite granularity of PC2 and PC3 pads. The positive directions of
$D_{\phi}^+$ and $D_{\phi}^-$ are indicated by the arrow. A $\pm{2\sigma}$
cut on these variables is illustrated by the box region inside the
dashed lines.
    }
\end{figure}

Despite being incorrectly reconstructed with large $\pt$, the
majority of the background particles have low momenta. While
travelling from the DC to the PC2 and PC3, they multiple scatter
and receive an additional deflection from the fringe field. This
causes a correlated deflection between the measured positions at
PC2, PC3, and the projections calculated from tracks measured by
the DC and PC1. The displacements in r-$\phi$ and $z$ directions
are represented by $D_{\phi}$ and $D_{z}$. Since the residual bend
depends on the $z$ component of the fringe field, which decreases
rapidly at large $|\eta|$, a fiducial cut of $|\eta|<$0.18 was
applied to ensure that the residual bend due to the fringe field
is almost independent of $z$. We focus on the displacement in
r-$\phi$, $D_{\phi}$, which are large for low momentum tracks due
to the residual bend. The $D_{\phi}$'s at PC2 and PC3 are
correlated with each other, as shown in Figure~\ref{fig:pc2pc3}.
Most of the tracks lie in a narrow window around the diagonal
line. The width of this window is given by the PC2 and PC3
detector resolutions, which are of the order of a few millimeters.
Multiple scattering and residual bend  broaden the matching
distribution along the diagonal line. To optimize background
rejection, we define two orthogonal projections,
\begin{eqnarray}
    \brf &=& \frac{1}{\sqrt{2}}(D_{\phi}^{pc2}+D_{\phi}^{pc3})\\
    \naf &=& \frac{1}{\sqrt{2}}(D_{\phi}^{pc2}-D_{\phi}^{pc3})\nonumber
 \label{eq:3}
\end{eqnarray}
$\brf$ is the variable along the correlated direction, $\naf$ is
the direction normal to $\brf$. A $\pm$2$\sigma$ cut on these
variables is applied in the data analysis. In the remaining
discussion, unless stated otherwise, only tracks satisfying these
cuts are included.
%
%

After matching cuts, the background level is less than 6\% for
$\pt<4$~GeV/$c$, but increases rapidly at higher $\pt$. For
$4<\pt<10$ GeV/$c$, the most significant remaining background
sources are $e^{+}e^{-}$ from conversion of photons close to the
DC and particles from weak decays of long lived particles, mainly
of $K^{\pm}$ and $K^{0}_{L}$. These backgrounds are estimated and
subtracted separately from the $\brf$ distribution for all tracks,
as will be discussed in the rest of this section.

To separate the two background sources, we take advantage of the
RICH to tag electrons. Charged particles with velocities above the
Cherenkov threshold $\gamma_{th} = 35$ ($CO_{2}$ radiator) will
emit Cherenkov photons, which are detected by photon multiplier
tubes (PMT) in the RICH~\cite{richnim}. We characterize the
Cherenkov photon yield for a charged particle by $\npmt$, the
number of PMTs with signals above threshold associated to the
track. For reconstructed electrons ($\pt> 150 MeV/c$), the average
number of associated PMTs is $\meanpmt \approx 4.5$. The
probability to find at least one PMT above threshold is more than
$ 99\%$. For pions, the Cherenkov threshold is 4.8 GeV/$c$, and
the number of associated PMTs reaches its asymptotic value only
well above 10 GeV/$c$; $\meanpmt$ increases from 1.4 at 6 GeV/$c$
to 2.8 at 8 GeV/$c$ and 3.6 at 10 GeV/$c$.

\begin{figure*}[t]
\includegraphics[width=0.75\linewidth]{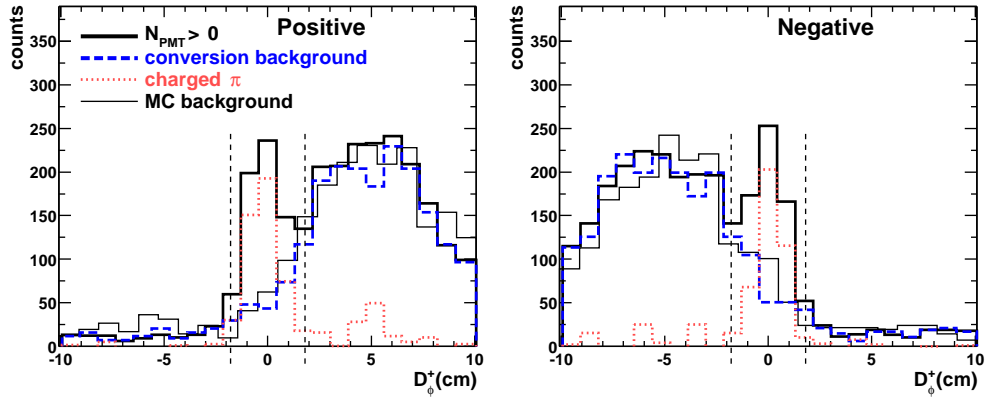}
\caption{\label{fig:bgsub1} (Color online) Background
contamination due to electrons, illustrated by the track match in
$D_{\phi}^+$ for tracks with associated RICH PMTs and $6 < \pt <
7$ GeV/$c$. The matching distributions are shown for minimum bias
events and separately for positive (left) and negative (right)
charged tracks. The first three distributions represent the raw
counts for all tracks with RICH association (thick solid line),
estimated conversion backgrounds (dashed line) and charged pions
(dot-dashed line) that were obtained by subtracting the dashed
line from the solid line. The thin solid line represents the
matching distribution of background electrons from Monte-carlo
simulation, arbitrarily scaled to match the data. The 2 $\sigma$
matching window are illustrated by the vertical dashed line. Since
$e^{+}$ and $e^{-}$ are deflected in opposite directions by the
fringe field, they are shifted to positive and negative
directions, respectively. }
\end{figure*}

Tracks ($N_{R}$) with at least one associated RICH PMT contain
both conversion electrons and real pions. Their matching
distributions in $\brf$ are presented in Figure~\ref{fig:bgsub1}
for a sample range of $6 < \pt < 7$ GeV/$c$. Also shown is the
matching distribution for conversion electrons from Monte-Carlo
simulation. The contributions from pions and electrons are clearly
distinguishable. For pions with $\pt<10$~GeV/$c$, $\meanpmt$ has
not reached its asymptotic value. A requirement of $\npmt\geq5$
rejects most of the pions while preserving a well-defined fraction
($R_{e}$) of the electrons. To measure $R_e$ from the data, we
select tracks with an apparent $\pt>10$~GeV/$c$.~\footnote{In this
$\pt$ range, the background yield decreases slowly with $\pt$,
while the true $\pi$ yield decreases rapidly as $\pt$ increases.
By comparing the measured $\piz$ spectrum from
PHENIX~\cite{ppg014} with the charged hadron spectrum before
background subtraction at $\pt>$ 10 GeV/$c$, the integrated signal
yield is estimated to be less than 3\% and thus may be neglected.}
The fraction of tracks with $\npmt\geq5$ is measured to be $R_{e}
= 0.458 \pm 0.05$. Both Monte-Carlo and data show a small
variation of $R_{e}$ with $\pt$ and centrality. This variation is
included in the error on $R_e$. The total electron background is
calculated using tracks with $\npmt\geq5$ ($N_e$) as, $N_e/R_e$.
The number of real pions in the RICH-associated sample for each
$\pt$ bin is then calculated as,
\begin{equation}
 S_R = N_R - \frac{N_e}{R_e} \qquad .
\end{equation}
With this method, a small fraction of genuine pions, which satisfy
$\npmt\geq5$, is subtracted. This fraction is negligible below 7
GeV/$c$, but increases rapidly towards higher $\pt$. This loss is
corrected using the PHENIX Monte-Carlo simulation. In this case, a
100\% error on the correction is assigned.
%
%

\begin{figure*}[t]
\includegraphics[width=0.75\linewidth]{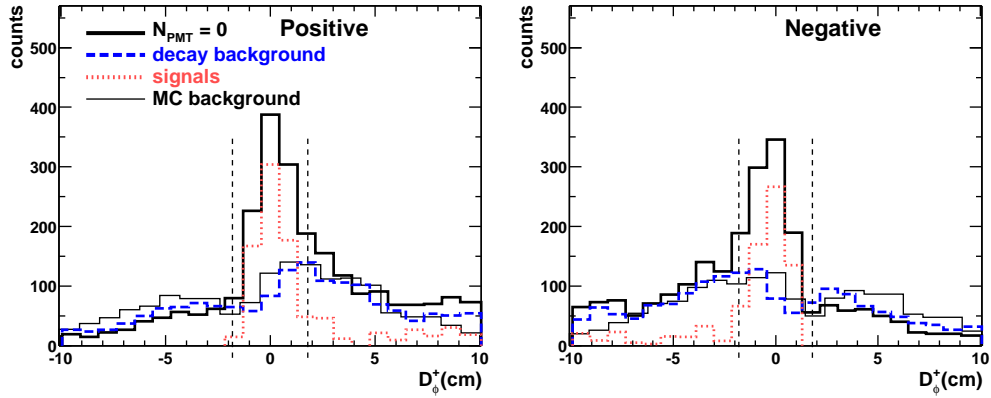}
\caption{\label{fig:bgsub2} (Color online) Background
contamination due to decays, illustrated by the track match in
$D_{\phi}^+$ for tracks without an associated RICH PMT and with
$6<\pt<7$ GeV/$c$, shown for minimum bias events and separately
for positive (left) and negative (right) charged tracks. The first
three distributions represent the raw counts for all tracks
without a RICH association (thick solid line), estimated decay
backgrounds (dashed line), and signal tracks (dot-dashed line)
that were calculated as the difference of the two. The thin solid
line represents the matching distribution of decay background from
Monte-carlo simulation, arbitrarily scaled to match the data. The
2 $\sigma$ matching window are illustrated by the vertical dashed
line. Outside the signal window, the shape of the dashed line
matches the solid line rather well, the difference of 10\% level
is taken into account in the error estimation of $R_{decay}$.}
\end{figure*}

The sample of tracks ($N_{NR}$) with no associated RICH PMT
contains a mixture of $\pi,K,p$, contaminated by the decay
background. Their matching distributions in $\brf$ are shown in
Figure~\ref{fig:bgsub2} for $6 < \pt < 7$ GeV/$c$, together with
the matching distribution for decay particles from MC simulation.
A Monte-Carlo study shows that the apparent momentum of these
tracks is nearly uncorrelated with true momentum and therefore the
distribution of this background in $\brf$ is nearly independent of
the apparent momentum. We select a nearly pure background sample
using tracks with reconstructed $\pt>10$~GeV/$c$ and measure the
ratio of the number of tracks passing a $|\brf|<2\sigma$ cut to
tracks in the interval $3\sigma <|\brf| <9\sigma$:
\begin{eqnarray}
R_{decay} &=& \frac{N_{NR}(p_T > 10 GeV/c, |D_{\phi}^+| < 2\sigma)}{N_{NR}(p_T > 10 GeV/c, 3\sigma <|D_{\phi}^+| <9\sigma)}\\
          &=& 0.424 \pm 0.05\nonumber
\label{eq:4}
\end{eqnarray}
The error quoted takes into account the small variation of
$R_{decay}$ with $\pt$ and centrality. Since the average yield of
real hadrons in this interval is small, we estimate the decay
contribution as a function of $\pt$ to be
$N_{NR}(3\sigma<|D_{\phi}^+|<9\sigma)\times R_{decay}$.
Finally, the signal is calculated as,
\begin{eqnarray}
    S_{NR} &=& N_{NR}(|D_{\phi}^+|<2\sigma) - \\
           & &N_{NR}(3\sigma<|D_{\phi}^+|<9\sigma)\times R_{decay}\nonumber
\end{eqnarray}
%
%

\begin{figure*}[t]
\includegraphics[width=0.75\linewidth]{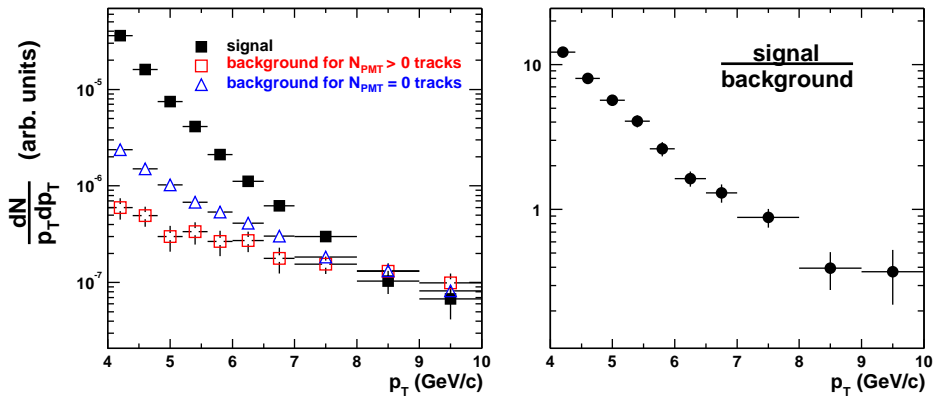}
      \caption{\label{fig:bgsub3}
(Color online) Amount of background estimated as function of $\pt$
for minimum bias collisions. The left figure shows the background
subtracted charged hadron spectra (filled square), the background
from $\gamma$ conversions (open square) and decays (open
triangle). The right figure shows the signal to background ratio.
Only statistical errors are shown.
 }
\end{figure*}

Figure~\ref{fig:bgsub3} gives the total signal, obtained as $S_R +
S_{NR}$, with the decay and conversion background subtracted. On
the right hand side, the signal-to-background ratio is shown. The
background increases with increasing $\pt$. At 4 GeV/$c$ the
signal-to-background ratio is about 10, and decreases to 1 at 7.5
GeV/$c$ and to $\sim$0.3 at 10 GeV/$c$.
%
%

Weak decays of short lived particles, mainly $K^{0}_{s}$,
$\Lambda$ and $\bar{\Lambda}$ within the magnetic field provide an
additional source of background. A significant fraction of this
background is subtracted using the $R_{decay}$ method described
above. However, secondary particles from decays close to the event
vertex are not subtracted since they are nearly indistinguishable
from primary particles. This ``feed-down'' contaminates the track
sample without the associated RICH PMTs, $S_{NR}$ (about 40\% of
all charged particles at high $\pt$), and needs to be subtracted
from the data.

To estimate the feed-down contribution we generate $\Au$ events
with HIJING~\cite{hijing}, reconstruct them through the PHENIX
Monte-Carlo simulation, and count the secondaries which survive
all analysis cuts. The secondaries from $\Lambda$ and
$\bar{\Lambda}$ decays are counted relative to the reconstructed
$(p + \bar{p})$, and correspondingly, those from $K^0_s$ relative
to $(K^+ + K^-)/2$. We tune the $(\Lambda + \bar{\Lambda})/(p +
\bar{p})$, $K^0_s/(0.5(K^+ + K^-))$ ratios by weighting the
particle distributions generated according to HIJING such that
they reproduce the nearly $\pt$-independent experimentally
observed ratios from $\Au$ collisions at $\snn$ = 130
GeV~\cite{ppg012,starkaon}.

The final feed-down contribution depends on the choice of the
$\Lambda$ and $K^0_s$ $\pt$ spectra and of their yields in the
high $\pt$ range where they are not measured. Both yields and
spectral shapes are varied within limits imposed by the spectrum
for tracks that do not fire the RICH. The average feed-down
contribution depends on $\pt$ and varies between 6 to 13\%
relative to the total charged hadron yield; it is subtracted from
the charged spectra. The systematic uncertainties are estimated
from the spread of the feed-down contributions obtained with
different assumptions. The uncertainties are approximately 60\% of
the subtraction, and depend on $\pt$ and centrality.

Table~\ref{tab:sys1} summarizes the systematic errors on the
background subtraction~\footnote{We should emphasize that, in the remaining discussion
unless stated otherwise, all systematic errors listed in Tables have been adjusted to $1\sigma$ errors.}.
All errors are correlated with $\pt$ and
are presented as relative errors to the charged hadron yield. The
uncertainty of the pion oversubtraction correction ($\delta_{\pi{loss}}$) 
was re-scaled
by the fraction of signal tracks with RICH association, i.e.
$S_R/(S_R + S_{NR})$. Errors on the scaling factors $R_{e}$ and
$R_{decay}$ were individually folded with the signal-to-background
ratios in the two samples. The resulting uncertainties on the
charged yields were then added in quadrature
($\delta_{R_{e}{\oplus}R_{decay}}$). The uncertainty of the
$K^0_s$, $\Lambda$, and $\bar{\Lambda}$ feed-down subtraction is
denoted by $\delta_{feeddown}$.

\begin{table*}
\caption{\label{tab:sys1}
Systematic errors on background
subtraction. All errors are given in percent and are quoted as
1$\sigma$ errors.  These errors are correlated with $\pt$.}
\begin{ruledtabular} \begin{tabular}{lllll}
$p_T$ (GeV/$c$) & $\delta_{\pi{loss}}$ (\%)  &
$\delta_{R_{e}{\oplus}R_{decay}}$(\%) & $\delta_{feeddown}$ (\%) &
total(\%) \\ \hline
$ < 5$        &             0.3            &            0.3         &                    5                        & 5 \\
5 - 6         &             0.6            &            1.8         &                    5                        & 5.3 \\
6 - 7         &             1.4            &            4.1         &                    8.5                      & 9.5 \\
7 - 8         &             4.6            &            7.1         &                    7.8                      & 11.5\\
8 - 9         &             9.9            &            17.6        &                    6                        & 21.1  \\
9 - 10        &             19.4           &            23.5        &                    6                        & 31.1  \\
\end{tabular}   \end{ruledtabular}
\end{table*}

\subsection{Corrections and systematic uncertainties}
\label{sec:Scorr}

\begin{figure}[t]
\includegraphics[width=1.0\linewidth]{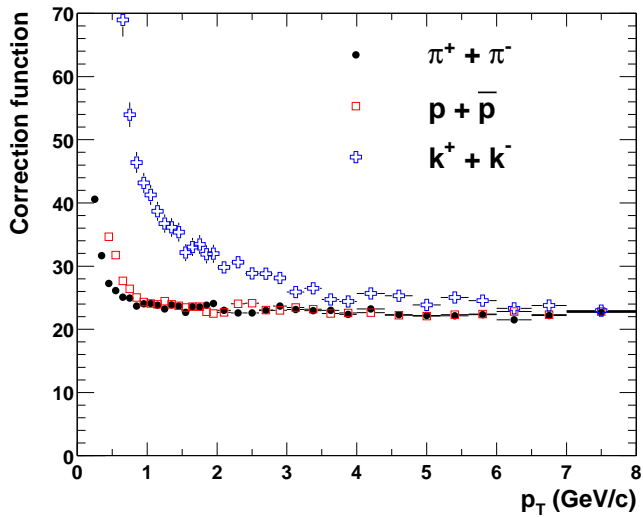}
     \caption{\label{fig:pidcorr}
(Color online) Averaged correction functions for $\pi^+$ and
$\pi^-$, $p$ and $\bar{p}$, and $K^+$ and $K^-$.
 }
\end{figure}

\begin{figure*}[t]
\includegraphics[width=0.75\linewidth]{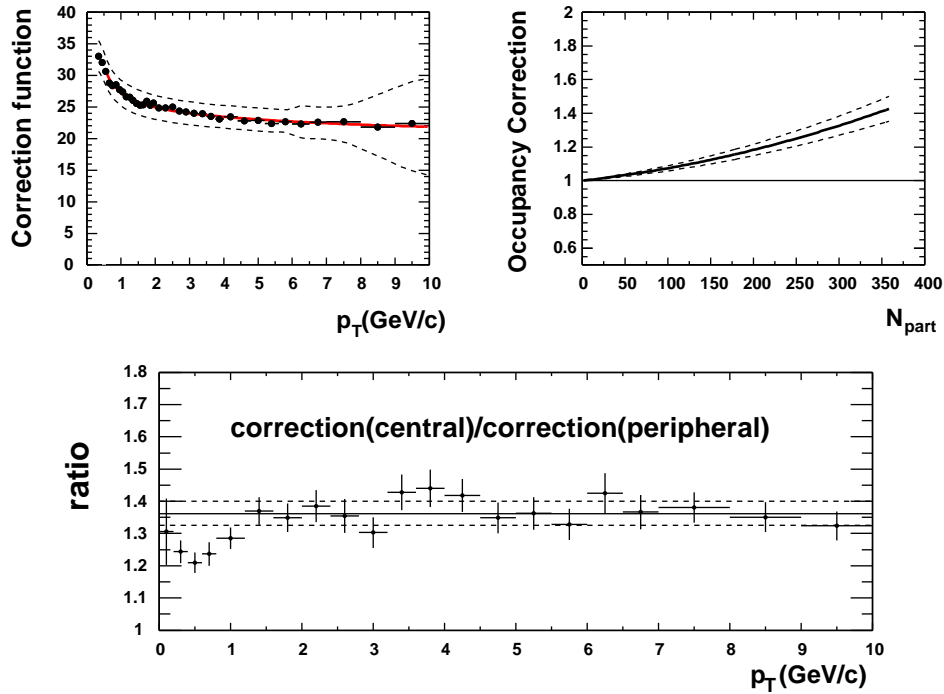}
     \caption{\label{fig:corr}
    (Color online) Functions used to correct the charged particle $\pt$ spectra. The upper
    left panel shows the $\pt$ dependent
    correction, $c(p_T)$. The upper right panel shows the centrality dependent correction,
        $c(N_{part})$. Systematic uncertainties are
        indicated by the dashed lines. The two corrections factorize at $\pt>1.5$~GeV/$c$, so
        that for given centrality the full correction function is given by
    $c(p_T) \times c(N_{part})$. The accuracy of this factorization
    is demonstrated in the lower panel. The ratio of the full
        correction for central collisions (5\% most central) to the correction
        for single particle events
    varies by less than $3$\% above 1.5 GeV/$c$ (the error bar is the statistical error from the Monte-Carlo calculation).
   }
\end{figure*}

After background subtraction, we have determined a single,
$\pt$-dependent correction function to correct the hadron spectra
for acceptance, decay in flight, reconstruction efficiency and
momentum resolution. This function is determined using a
GEANT~\cite{geant} Monte-Carlo simulation~\cite{phenixoff} of the
PHENIX detector in which simulated single tracks are reconstructed
using the same analysis chain applied to the real data. Because of
decays and multiple scattering, the correction function depends on
the particle species. This is reflected in
Figure~\ref{fig:pidcorr}, where the correction functions averaged
between $\pi^+$ and $\pi^-$, $K^+$ and $K^-$, $p^+$ and $p^-$ are
shown separately. At $\pt <3$~GeV/$c$, the kaon correction
function is significantly larger than those for pions and protons;
For $\pt>3$ GeV/$c$, this difference is less than 15\%. To take
into account this species dependence, we determine the correction
function separately for pions, kaons, protons, and their
anti-particles. The final correction function is then obtained by
combining the correction functions for the different particle
species weighted by the measured $\pt$-dependent particle
composition from~\cite{ppg026}. Above 2 GeV/$c$, where kaon data
are not available, we assume the $K/\pi$ ratio is constant within
$\pm10\%$ from the value observed at 2 GeV/$c$. This assumption
leads to a 2.5\% systematic error in the correction function. The
resulting correction function is plotted in the upper left panel
of Figure~\ref{fig:corr}. The sharp rise below 2 GeV/$c$ is due to
loss in acceptance and decays in flight. Above 2 GeV/$c$, the
correction decreases only slowly with $\pt$. For $\pt > 4$
GeV/$c$, the correction varies by less than $\pm$5\%.

The data are also corrected for efficiency losses due to detector
occupancy. Though this is negligible for peripheral collisions,
these losses are important in central collisions, and are
evaluated by embedding simulated tracks into real events. The
average track reconstruction efficiency in the active detector
area is larger than 98\% for peripheral collisions, but decreases
to $70\pm3.5\%$ for central collisions. As shown in the lower part
of Figure~\ref{fig:corr}, the efficiency loss is independent of
$\pt$ within a $\pm$3\% systematic uncertainty from 1.5 to 10
GeV/$c$. Based on this observation, the full correction can be
factorized into centrality-dependent (i.e. detector occupancy
dependent) correction function, $c(\npart)$, and $\pt$-dependent
correction function, $c(\pt)$. The centrality-dependent correction
function is shown on the upper right panel of
Figure~\ref{fig:corr}. Most of the efficiency loss is due to hit
overlaps, which can shift the hit positions in the DC or PC's
outside of the matching windows. The $\pm$2$\sigma$ matching
windows are larger at low $\pt$ to account for multiple scattering
(see Eq.~\ref{eq:2}), thus the tracks are less vulnerable to the
effect of hit merging. This effect has been taken into account by
applying a slightly smaller, $\pt$-dependent, occupancy correction
at $\pt < 1.5$ GeV/$c$.
%
%

Figure~\ref{fig:corr} also shows systematic errors on the correction
functions. These errors include not only the errors on the
correction itself, but also the uncertainty due to the background
subtraction procedure.

Finally, the inclusive charged hadron yield are obtained by
multiplying the $\pt$-dependent correction function, $c(\pt)$, and
centrality-dependent correction function, $c(\npart)$, with the
background subtracted spectra and dividing by the number of events
for every centrality class as:
\begin{eqnarray}
\frac{1}{N_{evts}}\frac{dN}{2\pi\pt d\pt d\eta} &=&
\frac{1}{N_{evts}}\times c(\pt)\times c(\npart)\\
& &\times \left(\frac{dN}{\pt d\pt d\eta}\right)^{bgr-subtracted}\nonumber
\label{eq:chyield}
\end{eqnarray}

The systematic errors on the spectra, which are common to all
centrality classes, are listed in Table~\ref{tab:sys2}. Sources of
systematic uncertainties are: the matching cuts
($\delta_{match}$), normalization ($\delta_{norm}$), particle
composition ($\delta_{mix}$), momentum resolution
($\delta_{reso}$), momentum scale ($\delta_{scale}$), and
background subtraction ($\delta_{bgr}$) from Table~\ref{tab:sys1}.
The normalization error is independent of $\pt$. All other errors
vary with $\pt$ but are highly correlated bin-to-bin, which means
that points in neighboring $\pt$ bins can move in the same
direction by similar factors.

\begin{table*}
\caption{\label{tab:sys2}
Systematic errors on the $\pt$ spectra.
All errors are given in percent and are quoted as 1$\sigma$
errors. They are either normalization errors or are $\pt$
correlated errors.}
  \begin{ruledtabular} \begin{tabular}{llllllll}
$\pt$ (GeV/$c$)& $\delta_{match}$(\%) & $\delta_{norm}$(\%)&
$\delta_{mix}$(\%)& $\delta_{reso}$(\%) &
$\delta_{scale}$(\%)&$\delta_{bgr}$(\%) & total(\%) \\ \hline
 $ < 1$     &   3.5              &  3.2          &              2.4           &          0.6        & 0.6                &    5         &  7.3 \\
 1 - 5      &   3                &  3.2          &              2.4           &          0.6        & 3                  &    5         &  7.6 \\
 5 - 6      &   3                &  3.2          &              1.8           &          0.6        & 3.6                &    5.3       &  7.9 \\
 6 - 7      &   3                &  3.2          &              1.8           &          0.6        & 3.3                &    9.5       &  11.1 \\
 7 - 8      &   3                &  3.2          &              1.8           &          0.6        & 3.1                &    11.5      &  12.8 \\
 8 - 9      &   3                &  3.2          &              1.8           &          0.9        & 3.1                &    21.1      &  21.9 \\
 9 - 10     &   3                &  3.2          &              1.8           &          5.3        & 3.1                &    31.1      &  32.1 \\
\end{tabular}   \end{ruledtabular}
\end{table*}

The centrality-dependent systematic errors are quantified in terms
of the central-to-peripheral ratio, $R_{cp}$, as given in
Table~\ref{tab:sys3}. Besides the uncertainty on the occupancy
correction ($\delta_{occupancy}$) illustrated in
Figure~\ref{fig:corr}, the background subtraction procedure has a
centrality-dependent uncertainty. As discussed in
Section~\ref{sec:Sbg}, the errors on $R_e$ and $R_{decay}$ reflect
the $\pt$ and centrality dependencies. The centrality-dependent
part contributes about half of the error on both $R_e$ and
$R_{decay}$, and hence does not cancel in $R_{cp}$. Since the
errors on  $R_e$ and $R_{decay}$ are independent, the uncertainty
on $R_{cp}$ is approximately equal to
$\delta_{R_{e}{\oplus}R_{decay}}$ from Table~\ref{tab:sys1}.
Finally, $\delta_{feeddown}$ is the centrality-dependent error
from feed-down subtraction.

\begin{table*} \caption{\label{tab:sys3}
Systematic errors on the central-to-peripheral ratio. All errors
are given in percent and are quoted as 1$\sigma$ errors. Most of the
systematic errors listed in Table~\ref{tab:sys2} cancel in the
central-to-peripheral ratio. Only those errors that are
uncorrelated with centrality are shown here.}
  \begin{ruledtabular} \begin{tabular}{lllll}
$\pt$ (GeV/$c$) & $\delta_{occupancy}$(\%) &
$\delta_{feeddown}$(\%)&  $\delta_{R_{e}{\oplus}R_{decay}}$ (\%) &
total(\%) \\\hline
$ < 6$        &     5              &    5               &    1.8                                 & 7.3        \\
6 - 7         &     5              &    5               &    4.1                                 & 8.2        \\
7 - 8         &     5              &    5               &    7.1                                 & 10         \\
8 - 9         &     5              &    5               &    17.6                                & 19         \\
9 - 10        &     5              &    5               &    23.5                                & 24.6       \\
\end{tabular}    \end{ruledtabular}
\end{table*}


\section{RESULTS}
\label{sec:Sresults}

\subsection{Inclusive charged hadron $\pt$ spectra} \label{sec:Scharge}

\begin{figure*}[t]
\includegraphics[width=0.7\linewidth]{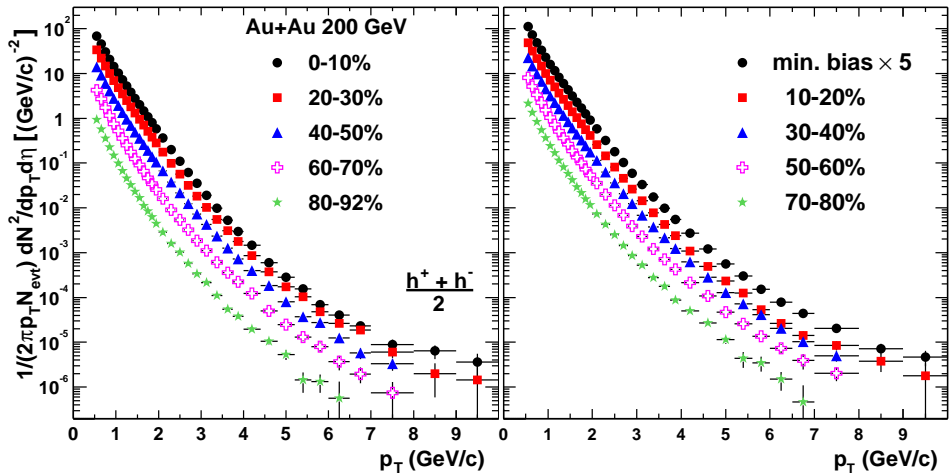}
     \caption{\label{fig:spec1}
    (Color online) $\pt$ spectra of charged hadrons for minimum bias collisions along with
spectra for 9 centrality classes derived from the pseudo-rapidity
region $|\eta|<0.18$. The minimum bias spectrum has been
multiplied by 5 for visibility. Only statistical errors are shown
in the spectra. Most of the $\pt$ dependent systematic errors are
independent of centrality and are tabulated in
Table~\ref{tab:sys2}.
   }
\end{figure*}

\begin{figure*}[t]
\includegraphics[width=0.7\linewidth]{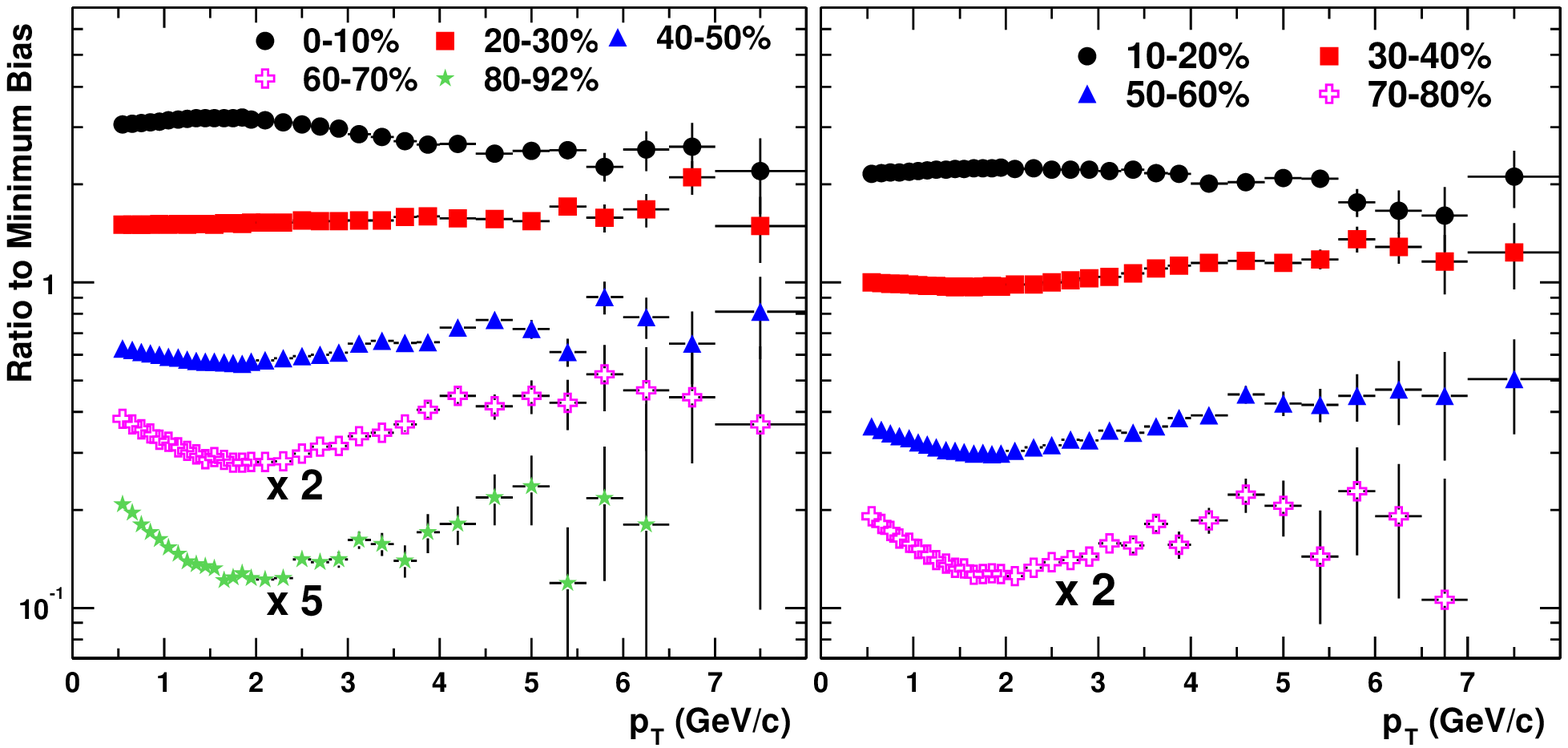}
     \caption{\label{fig:spec2}
     (Color online) Ratios of centrality selected $\pt$ spectra to the minimum bias
     spectrum. Ratios for peripheral classes are scaled up for
     clarity. For the $\pt$ range shown, most of the systematic
     errors cancel in the ratio. The remaining systematic errors
     that can change the shape are less than 10\% (see Table~\ref{tab:sys3})
    and are correlated
     bin-to-bin in $\pt$.
   }
\end{figure*}

Figure~\ref{fig:spec1} shows the inclusive charged hadron $\pt$
spectra for various centrality classes. All spectra exhibit
power-law tails at high $\pt$. But for peripheral collisions, the
power-law shape is more concave than for central collisions. More
details of the centrality dependence of the spectral shape can be
seen from Figure~\ref{fig:spec2}, which shows for each centrality
class the ratio of the spectra to the minimum-bias spectrum. In
these ratios, most systematic errors cancel or affect the overall
scale only. The characteristic centrality dependence of the shape
already observed in $\snn$ = 130~GeV $\Au$
collisions~\cite{star,ppg013} is more apparent at $\snn$ =
200~GeV. In peripheral collisions, the ratio decreases up to $\pt
\sim 2$ GeV/$c$ and then rises up to about 4 GeV/$c$. The trends
are reversed in the most central collisions. In the range above
4--5 GeV/$c$, all ratios appear to be constant as function of
$\pt$, which would imply that they have a similar centrality
independent shape.
%
%
%
%

\begin{figure}[t]
\includegraphics[width=1.0\linewidth]{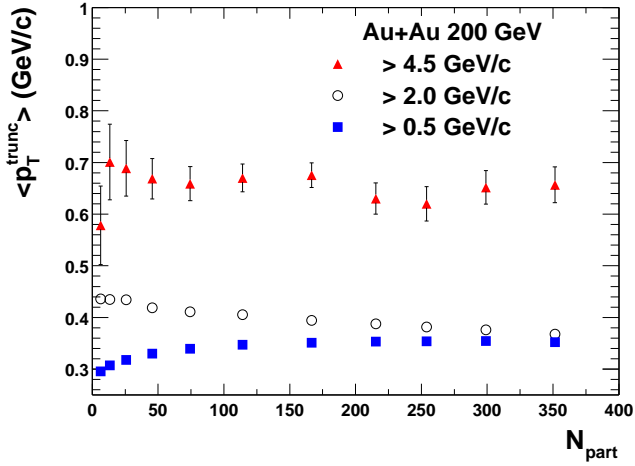}
    \caption{\label{fig:trunpt}
    (Color online) Centrality dependence of $\trun$, the
    average $\pt$ of charged particles above a $\pt$
    threshold as defined in Eq.~\ref{eq:5}. Shown
    are $\trun$ values for three $p_T^{min}$ cuts, with $p_T^{min} =$ 0.5, 2 and
    4.5~GeV/$c$ respectively. The errors shown are statistical
    only. The systematical errors for all data points are less than 3\%.
    }
\end{figure}

Based on the different trends observed in Figure~\ref{fig:spec2},
we can distinguish three $\pt$ regions: 0.5--2, 2--4.5 and $> 4.5$
GeV/$c$. The different centrality dependence of the spectral shape
in these regions can be quantified by a truncated average $\pt$:
\begin{equation}
     \langle p_T^{trunc} \rangle \equiv
     \frac{\int_{p_{T}^{min}}^{8~GeV/c} p_{T} \cdot dN/dp_{T}}
     {\int_{p_{T}^{min}}^{8~GeV/c} dN/dp_{T}} - p_{T}^{min},
     \label{eq:5}
\end{equation}
which is insensitive to the normalization of the spectra. The
upper bound of 8 GeV/$c$ in the integral is given by the limited
$\pt$ reach for peripheral centrality classes as shown in
Figure~\ref{fig:spec1}. In Figure~\ref{fig:trunpt}, the values of
$\trun$ for the three $p_T^{min}$ values are plotted as function
of centrality, represented by the average number of participating
nucleons ($\npart$) for each centrality class.
%
%

For $p_T^{min} = 0.5$~GeV/$c$, where particle production is
expected to be governed by soft physics, $\trun$ increases with
$\npart$. This trend is also seen for the average $\pt$ of
identified charged hadrons, and reflects the increased radial flow
of soft particles in more central collisions~\cite{ppg026}. For
$p_T^{min} = 2$~GeV/$c$, the trend is significantly different. For
peripheral collisions, $\trun$ is substantially larger than the
value obtained with $p_T^{min}$ = 0.5~GeV/$c$ due to the presence
of the power-law tail. With increasing $\npart$, $\trun$ for
$p_T^{min}$ = 2 GeV/$c$ decreases and the values obtained with
$p_T^{min} = 0.5$ and 2~GeV/$c$ approach each other, which
indicates an almost exponential spectrum in central collisions
between 0.5 and 2 GeV/$c$. For the highest $\pt$ range ($p_T^{min}
= 4.5$~GeV/$c$), $\trun$ is approximately constant. This implies
that the shape of the spectrum is nearly independent of
centrality, as would be expected if this region is dominated by
hard scattering.

\begin{figure}[t]
\includegraphics[width=1.0\linewidth]{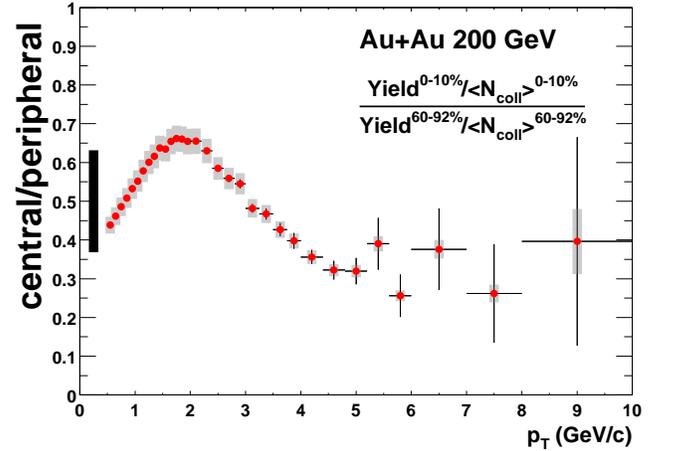}
    \caption{\label{fig:cp}
    (Color online) Ratio of charged hadron yields per nucleon-nucleon collision
    between central (0-10\%) and  peripheral (60-92\%) $\Au$ collisions.
    The solid error bars
    on each data point are statistical. The error bar on the left
    hand side of the figure is the overall scale error relative to
    0.5, which is the quadrature sum of (i) the uncertainty of
    $\ancoll$ (see Table~\ref{tab:sys0}) and (ii) the uncertainty on
    the occupancy correction ($\delta_{occupancy}$). The shaded error
    band on each data point is the $\pt$ dependent systematic error from
    $\delta_{R_{e}{\oplus}R_{decay}}$ and centrality
    dependent feed down correction ($\delta_{feeddown}$) as given in
    Table~\ref{tab:sys3}. }
\end{figure}

However, the yields at high $\pt$ do not scale with the number of
nucleon-nucleon collisions; they are suppressed comparing to the
binary collision scaling expected for hard scattering processes.
This can be clearly seen from Figure~\ref{fig:cp}, which shows
$R_{cp}$, the ratio of yields for central and peripheral
collisions normalized to the average number of nucleon-nucleon
collisions in each event sample. The ratio is below unity for all
$\pt$. The three $\pt$ regions show different trends as outlined
in the discussion of Figure~\ref{fig:trunpt}: (i) In the ``soft''
region with $\pt < 2$~GeV/$c$, the ratio increases as function of
$\pt$. (ii) In the ``hard'' region with $\pt > 4.5$~GeV/$c$, the
suppression appears to be constant at $\sim 0.3$, which again
indicates that the spectra have a similar shape, but with the
yield in central collisions being suppressed by a constant factor
from 4.5 to ~10~GeV/$c$. (iii) In the transition region from 2 to
$\sim$4.5~GeV/$c$, the ratio decreases as function of $\pt$.
%
%

\subsection{Suppression of high $\pt$ hadrons in $\Au$ at $\snn$ = 200~GeV} \label{sec:Scp_pi0}

At finite $Q^2$, nuclear modifications of the parton
distribution~\cite{EKS98,dima} and initial~\cite{cronin} and final
state~\cite{Gyu90} interactions of the scattering partons can
modify the high-$\pt$ hadron production rates in hard scattering
processes. Medium modifications of hadron spectra are often
quantified by the ``nuclear modification factor'' $R_{AA}$, which
we calculate for each centrality class as the ratio of the yield
per nucleon-nucleon collision in $\Au$ to the yield in
nucleon-nucleon collisions:
\begin{equation}
R_{AA}(p_T,\eta) =  \left(\frac{1}{N_{evt}}
\frac{d^{2}N^{A+A}}{dp_T d\eta}\right)  / \left(\frac{\langle
N_{coll} \rangle}{\sigma^{N+N}_{inel}}
\frac{d^{2}\sigma^{N+N}}{dp_T d\eta}\right) \label{eq:6}
\end{equation}
$\ancoll/\sigma^{N+N}_{inel}$ is the average Glauber
nuclear overlap function, $\taa$, for each centrality class. In
order to calculate $R_{AA}$, we need a reference spectrum for
nucleon-nucleon collisions. Due to the lack of charged hadron data
with sufficient reach in $\pt$ from our own experiment, we
construct the $\nn$ reference for charged hadrons from the $\piz$
spectra in $\pp$ collisions at $\sqrt{s} = 200$ GeV/$c$ measured
by PHENIX~\cite{ppg024}, and the charged hadron to pion ratio
observed in other experiments, as described below.

The PHENIX $\piz$ spectra from $\pp$ collisions are measured out
to 14~GeV/$c$. These data can be parameterized by a power-law
function,
\begin{equation}
\label{eq:7} \frac{1}{2\pi\pt}\frac{d^{2}\sigma^{\piz}_{N+N}}{
dp_T d\eta} = A{\left(\frac{p_0}{p_0 + p_T}\right)}^n \qquad ,
\end{equation}
with A = 386~mb/(GeV/$c$)$^2$, $p_0$ = 1.219 GeV/$c$, and n =
9.99~\cite{ppg024}.

In $\pp$ experiments at the ISR, the $h/\pi$ ratio was measured to
be $1.6\pm0.16$, independent of $\pt$ from 1.5 to 5~GeV/$c$, and
independent of $\sqrt{s}$ from 23 to 63~GeV~\cite{ISR}. Below 1.5
GeV/$c$, $h/\pi$ decreases towards lower $\pt$. The ISR data are
consistent with data on $\pi,K,P$ production from FNAL E735
experiment~\cite{E735} at $\sqrt{s}$ = 1.8 TeV. The $h/\pi$ ratio
computed from these data increases with $\pt$ and reaches a value
of 1.6 at the end of the measured $\pt$ range, $\sim$1.5~GeV/$c$.
At high momentum, a $h/\pi$ ratio of $\sim1.6$ is also observed
for quark and gluon jet fragmentation in $e^+e^-$ collisions at
LEP by the DELPHI Collaboration~\cite{delphi}. Finally, charged
hadron data measured by PHENIX in $\pp$ collisions and data
measured by UA1~\cite{UA1} in $\bar{p}+p$ collisions, both at
$\sqrt{s}$ = 200~GeV/$c$, give consistent $h/\pi$ ratios when
compared to the PHENIX $\pp$ $\piz$ data.

Based on these findings, we assume that $h/\pi$ is constant above
1.5~GeV/$c$ in $\pp$ collisions at RHIC and that we can scale up
the $\piz$ cross-section (Eq.~\ref{eq:7}) by this factor to obtain
a reference for charged hadron production. To be consistent with
the data described above, we correct this reference below 1.5
GeV/$c$ using an empirical function,
\begin{equation}
\label{eq:8}
    r(\pt) = \left\{\begin{array}{ll}
    R_{h/\pi} - a{\left(p_{max} - \pt\right)}^2 &\textrm{ for $p_T \le p_{max}$}
    \\
    R_{h/\pi} &\textrm{  for $p_T > p_{max}$}
    \end{array}\right.\qquad ,
\end{equation}
where $R_{h/\pi}$ = 1.6, $p_{max} $ = 1.6 GeV/$c$ and a = 0.28
(GeV/$c$)$^{-2}$. The charged hadron reference used in this
analysis is then given by the product of the power-law function
from Eq.~\ref{eq:7} and the empirical correction from
Eq.~\ref{eq:8} as:
\begin{equation}
\label{eq:9} \frac{1}{2\pi\pt}\frac{d^{2}\sigma^{h^+ +
h^-}_{N+N}}{dp_T d\eta} = A{\left(\frac{p_0}{p_0 + p_T}\right)}^n
\times r(\pt) \qquad .
\end{equation}

The systematic errors on the charged hadron $\nn$ reference are
summarized in Table~\ref{tab:sys4}. The main sources of
uncertainties include: (i) the systematic errors on the absolute
normalization of the PHENIX $\piz$ data ($\delta^{\piz}_{norm}$),
which are independent of $\pt$, (ii) uncertainties due to the
power-law fit to the $\piz$ data ($\delta^{\piz}_{fit}$), and
(iii) uncertainties on $R_{h/\pi}$ ($\delta_{h/\pi}$), which are
estimated from the spread of $R_{h/\pi}$ obtained from different
data sets used to constrain $h/\piz$.

\begin{table*}
\caption{\label{tab:sys4} Systematic errors on the charged hadron
$\nn$ reference spectrum. All errors are given in percent and are
quoted as 1$\sigma$ errors. Positive and negative errors are given
separately where appropriate. Most of the errors are correlated
with $\pt$.}
   \begin{ruledtabular}  \begin{tabular}{cccccc}
$\pt$ (GeV/$c$)\hspace{2mm} &\hspace{2mm}
$\delta^{\piz}_{norm}$(\%)\hspace{2mm}
&\hspace{2mm}$\delta^{\pi^0}_{fit}$(\%)\hspace{2mm} &\hspace{5mm}
$\delta_{R_{h/\pi}}$(\%) \hspace{7mm}&\hspace{10mm} total
(\%)\\\hline
0.75            &$\pm$ 10.4             & -3.9 + 9.1       & \hspace{5mm}-15.1 + 5.9                   & \hspace{10mm}-18.7 + 15.0\\
1.00            &$\pm$ 10.4             & -4.1 + 8.9       & \hspace{5mm}-14.4 + 5.9                   & \hspace{10mm}-18.3 + 14.9\\
1.50            &$\pm$ 10.4             & -4.6 + 8.3       & \hspace{5mm}-11.6 + 5.9                   & \hspace{10mm}-16.3 + 14.6\\
2.00            &$\pm$ 10.4             & -5.1 + 7.7       & \hspace{5mm} -7.9 + 5.9                   & \hspace{10mm}-14.0 + 14.2\\
2.50            &$\pm$ 10.4             & -5.5 + 7.2       & \hspace{5mm} -5.9 + 5.9                   & \hspace{10mm}-13.1 + 13.9\\
3.00            &$\pm$ 10.4             & -5.9 + 6.7       & \hspace{5mm} -5.9 + 5.9                   & \hspace{10mm}-13.3 + 13.7\\
3.50            &$\pm$ 10.4             & -6.4 + 6.4       & \hspace{5mm} -5.9 + 5.9                   & \hspace{10mm}-13.5 + 13.5\\
4.50            &$\pm$ 10.4             & -7.5 + 6.5       & \hspace{5mm} -5.9 + 5.9                   & \hspace{10mm}-14.1 + 13.6\\
5.50            &$\pm$ 10.4             & -8.9 + 7.9       & \hspace{5mm} -5.9 + 5.9                   & \hspace{10mm}-14.9 + 14.3\\
6.50            &$\pm$ 10.4             &-10.7 + 10.5      & \hspace{5mm} -5.9 + 5.9                   & \hspace{10mm}-16.0 + 15.9\\
7.50            &$\pm$ 10.4             &-12.9 + 14.3      & \hspace{5mm} -5.9 + 5.9                   & \hspace{10mm}-17.6 + 18.7\\
8.50            &$\pm$ 10.4             &-15.8 + 19.4      & \hspace{5mm} -5.9 + 5.9                   & \hspace{10mm}-19.8 + 22.8\\
9.50            &$\pm$ 10.4             &-19.3 + 25.9      & \hspace{5mm} -5.9 + 5.9                   & \hspace{10mm}-22.7 + 28.5\\
\end{tabular}   \end{ruledtabular}
\end{table*}

%
%
\begin{figure*}[ht]
\includegraphics[width=0.85\linewidth]{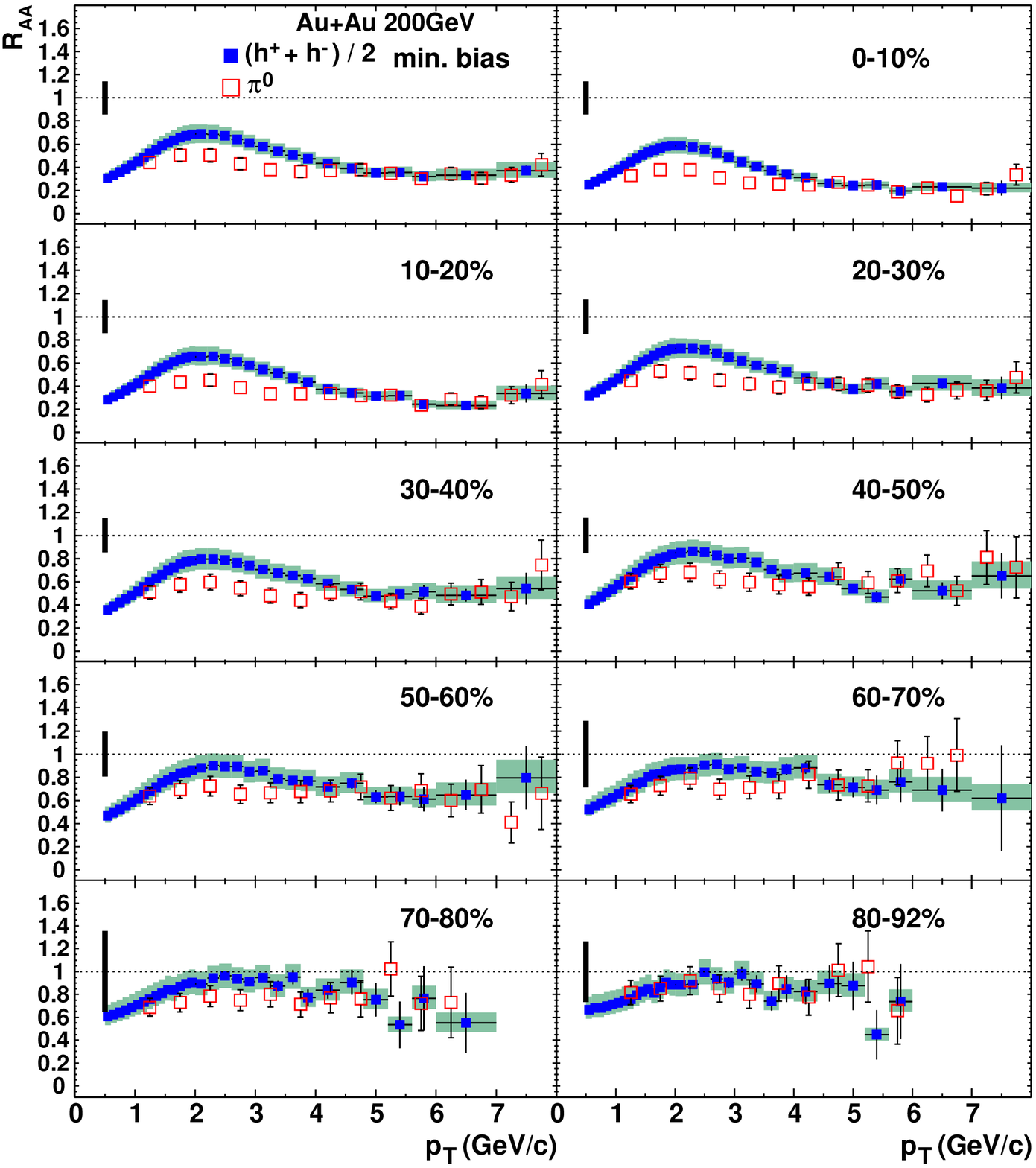}
     \caption{\label{fig:Raa}
     (Color online) $R_{AA}$ for $(h^{+} + h^{-})/2$ and $\pi^{0}$ as function of $p_{T}$
     for minimum bias and 9 centrality classes according to the ``{\it Fine}'' type of centrality
     classes defined in Table~\ref{tab:sys0}.
     The error bars on the $\pi^{0}$ data
     points include statistical and systematical errors on the $\Au$
     data and the $\nn$ reference. The error bars on $(h^{+} + h^{-})/2$ data points
     are statistical errors only. The common normalization errors ($\delta^{\piz}_{norm}$ from Table~\ref{tab:sys4}) on the
     references for charged hadrons and $\pi^0$s are added in
     quadrature with the uncertainty on $\ancoll$ and are indicated by
     the black bar on the left side of each panel. This error ranges
     from 15\% to 36\% from central to peripheral collisions and can
     shift all points in the charged and neutral pion $R_{AA}$ up and down
     together. The shaded band on charged $R_{AA}$ includes the
     remaining systematic errors on the charged $\nn$ reference summed in
     quadrature with the systematic errors from the $\Au$ spectra. This
     error amounts to -12.5\% + 18\% at low $p_T$ and changes to $\pm12.5$\%
     at $\pt$ = 4.5 GeV/$c$ and $\pm$18.5\% at $\pt = 8$ GeV/$c$.
   }
\end{figure*}

Figure~\ref{fig:Raa} shows the nuclear modification factor
$R_{AA}(p_T)$ for charged hadrons from minimum bias and nine
centrality classes. The systematic errors on $\raa$ are described
in the figure captions. At low $\pt$, the charged hadron $R_{AA}$
increase monotonically up to 2 GeV/$c$ for all centrality classes.
At $\pt>$ 2 GeV/$c$, $R_{AA}$ remains constant and close to unity
for the most peripheral centrality class. However, in central
collisions, it decreases at higher $p_T$, down to an approximately
constant value of 0.2--0.3 for $\pt>$ 4--5 GeV/$c$. This is
consistent with Figure~\ref{fig:cp}, where the central to
peripheral ratio also saturates above 4--5 GeV/$c$. This
approximately $\pt$ independent suppression pattern has been
interpreted as a result of the detailed interplay between the
Cronin effect, nuclear shadowing, and partonic energy
loss~\cite{ivan}.

\begin{figure*}[t]
\includegraphics[width=0.8\linewidth]{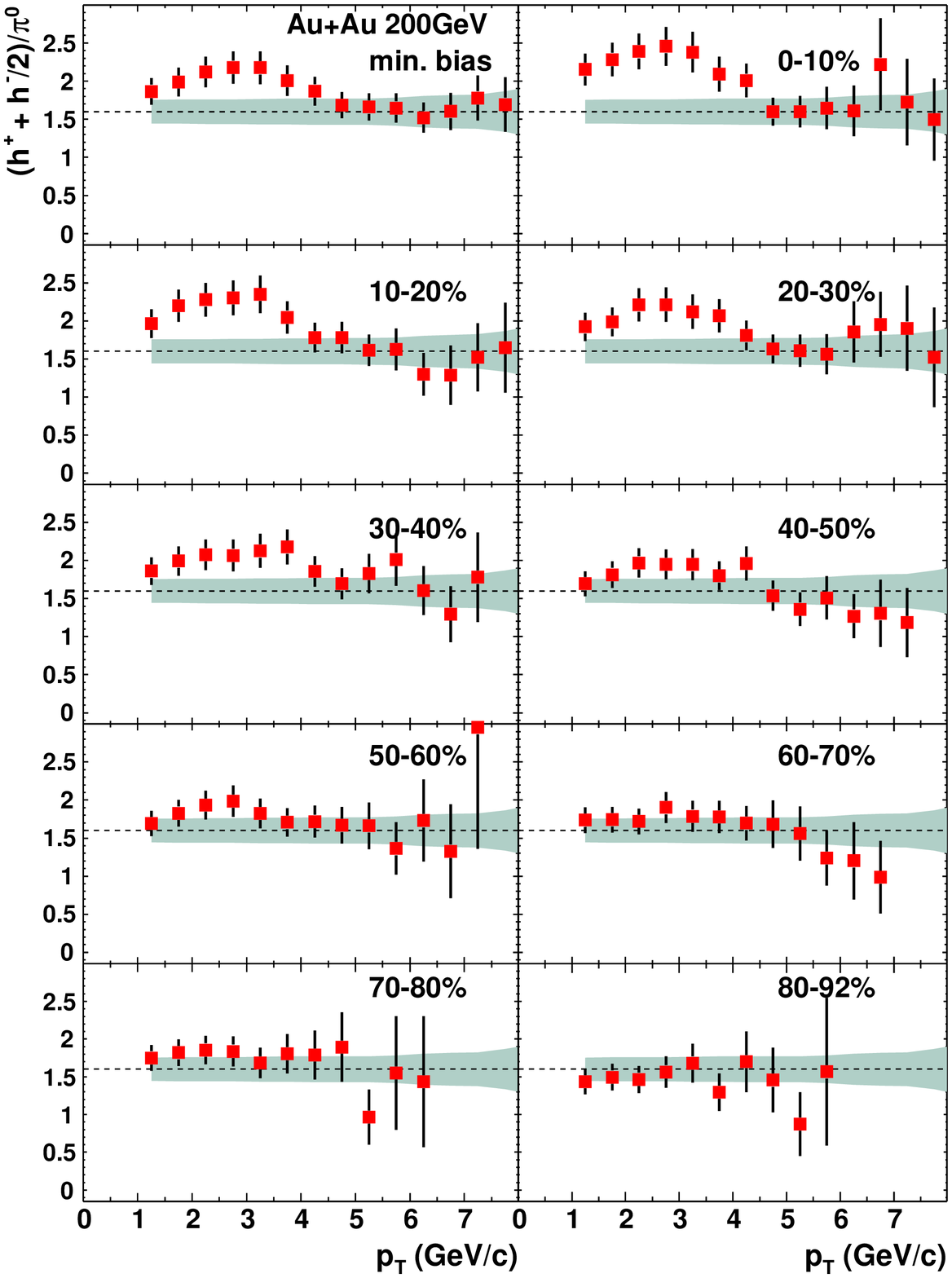}
     \caption{\label{fig:hpi}
     (Color online) Charged hadron to $\piz$ ratios for minimum bias events and 9 centrality
     classes according to the ``{\it Fine}'' type of centrality classes defined in
     Table~\ref{tab:sys0}.
     The error bars represent the quadratic sum of
     statistical and point-by-point systematic errors from $(h^++h^-)/2$
     and $\piz$. The  shaded band shows the percent normalization
     error (dominantly from $(h^++h^-)/2$ data) common to all centrality
     classes. The dashed line at 1.6 is the $h/\pi$ ratio measured in
     $\pp$~\cite{ISR} and $e^+e^-$~\cite{delphi} collisions.
     }
\end{figure*}

Also shown in Figure~\ref{fig:Raa} are $\raa$ for neutral pions
from ref.~\cite{ppg014}. The neutral pion $\raa$ values also seem
to reach maximum around 2 GeV/$c$, but the changes are smaller
than those for charged hadrons. Except for the most peripheral
bin, the neutral pion $\raa$ are always below the charged $\raa$
in the range of 2 $< p_T <$ 4.5 GeV/$c$. However, at $p_T > 4.5$
GeV/$c$, $\raa$ for both neutral pions and hadrons saturate at
roughly the same level, indicating a similar suppression for
neutral pions and charged hadrons at high $\pt$.

The fact that the neutral pion $\raa$ values are smaller than
inclusive charged hadron $\raa$ at intermediate $\pt$ ($ 2 < p_T <
4.5$ GeV/$c$) has already been observed at $\sqrt{s_{NN}} = 130$
GeV~\cite{ppg003}. This difference can be explained by the large
$p/\pi$ ratio observed in the same $p_T$ range in central $\Au$
collisions~\cite{ppg006,ppg015}. This large relative proton and
anti-proton yield indicates a deviation from the standard picture
of hadron production at $\pt>2$ GeV/$c$, which assumes that the
hadrons are created by the fragmentation of energetic partons.
Such a deviation has led to models of quark
coalescence~\cite{coalence} or baryon junctions~\cite{junction} as
the possible mechanisms to enhance the proton production rate at
medium $p_T$. Both models predict that baryon enhancement is
limited to $p_T <5$ GeV/$c$, beyond which jet fragmentation should
eventually become the dominant production mechanism for all
particle species. In that case, one would expect a similar
suppression factor for charged hadron and $\pi^0$, in agreement
with the data at $p_T
>4.5$ GeV/$c$. Recently, the difference of $\raa$ between charged hadrons and pions
was also argued as
the consequence of centrality and particle species dependent
$\langle k_T\rangle$ broadening effect~\cite{xiaofei}.

If hard-scattering dominates charged hadron production at
$\pt>4.5$ GeV/$c$, the particle composition should be determined
by the jet fragmentation function, similar to nucleon-nucleon
collisions. Figure~\ref{fig:hpi} shows $h/\piz$ for all centrality
classes. The systematic errors are explained in the figure
captions. In the most peripheral collisions, the $h/\piz$ ratio is
consistent with the $\pp$ values down to $\pt \approx 2$~GeV/$c$.
In central collisions, the $h/\piz$ ratio is enhanced by as much
as 50\% above the $\pp$ value in the region $1<\pt<4.5$~GeV/$c$.
This enhancement gradually decreases towards more peripheral
collisions and reflects the difference of $\raa$ between the
charged hadrons and $\piz$s, which is due to large baryon
contribution. The enhancement also strongly depends on $\pt$: It
reaches a maximum between 2.5 and 3.5 GeV/$c$, then decreases. At
$\pt>4.5$~GeV/$c$, the $h/\piz$ ratios for all centralities reach
an approximately constant value of 1.6, which is consistent with
the $h/\pi$ value observed in $\pp$~\cite{ISR} collisions and in
jet fragmentation in $e^+e^-$~\cite{delphi} collisions. The
similarity of the spectral shape and of the particle composition
between $\Au$ and $\pp$ collisions suggest that fragmentation of
hard-scattered partons is the dominant mechanism of particle
production in $\Au$ collisions above $\pt$ of 4--5 GeV/$c$,
regardless of the fact that the yields do not scale with $\ncoll$.

%
%
\begin{figure}[ht]
\includegraphics[width=1.0\linewidth]{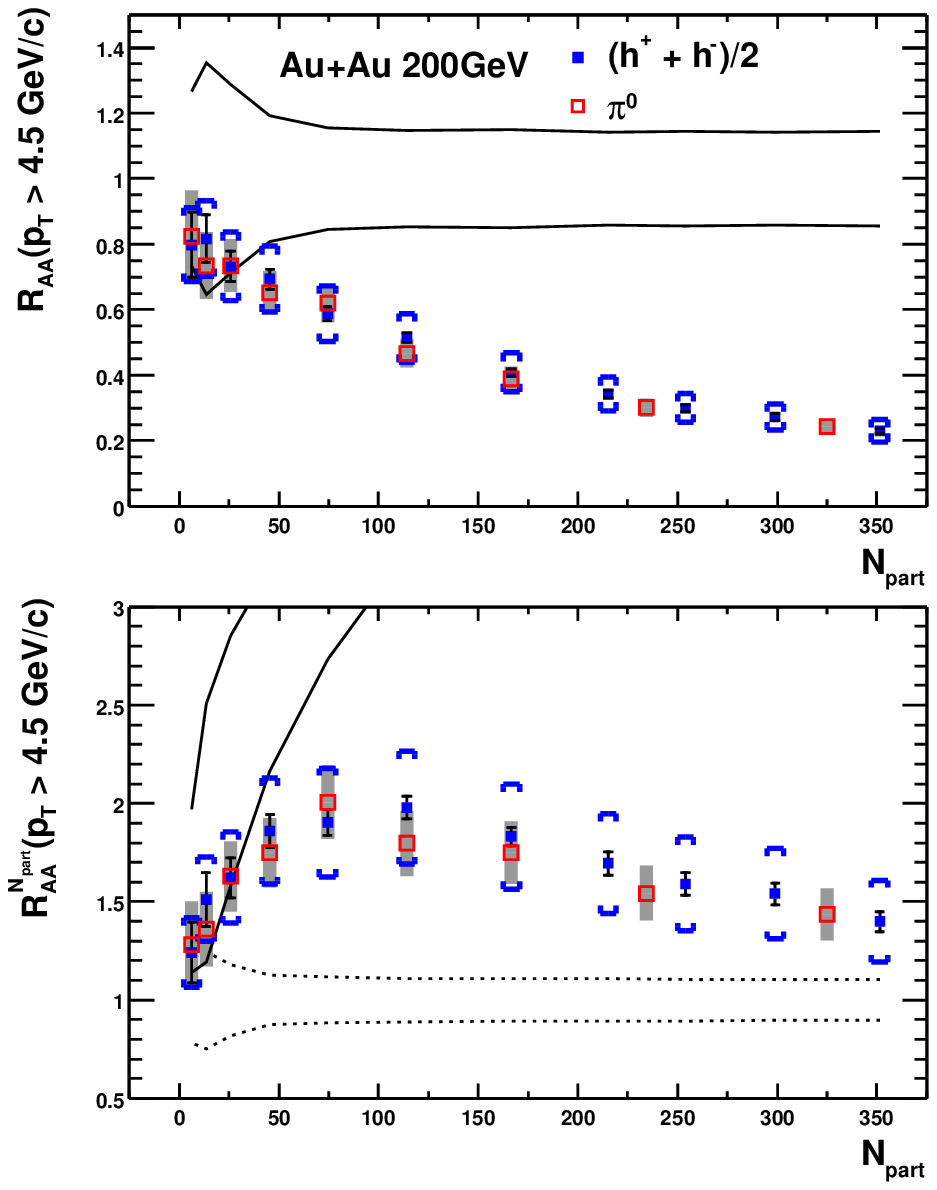}
\caption{\label{fig:Raainte} (Color online) $\Au$ yield integrated
for $\pt>4.5$ GeV/$c$ over the $\nn$ yield, normalized using
either $\ncoll$ ($\raa$ in the top panel) or $\npart$ ($\raap$ in
the bottom panel), plotted as function of $\anpart$. The bands
represent the expectation of binary collisions (solid) and
participant pair (dashed) scaling. The width of the bands gives
the systematic errors on $N_{coll}$ ($N_{part}$) added in
quadrature with the common normalization errors on the $\nn$
references for charged hadrons and neutral pions. For charged
hadrons, the statistical errors are given by the bars. The
systematic errors, which are not common with the errors for
neutral pions and which are correlated in $\pt$ are shown as
brackets. The shaded bars around each neutral pion points
represent the systematic and statistical errors, these errors are
not correlated with the errors shown for the charged hadron data.}
\end{figure}

Since $\raa$ values for charged hadrons and $\piz$s are
approximately constant at $\pt>4.5$~GeV/$c$, we can quantify the
centrality dependence of the $\raa$ value by calculating it from
yields integrated above 4.5 GeV/$c$. The upper panel of
Figure~\ref{fig:Raainte} shows $\raa$ for $\pt>4.5$ GeV/$c$ as
function of $\npart$. The $\raa$ values for charged hadrons and
$\piz$ agree for all centrality classes within errors. In
peripheral collisions with $\npart < 50$, $\raa$ is consistent
with binary collision scaling. With increasing $\npart$, $\raa$
decreases monotonically, reaching a value of $0.23\pm0.03$ (0-5\%
most central) for charged hadrons and $0.24\pm0.02$ (0-10\% most
central) for $\piz$s. There is an additional 14\% error common to
charged hadrons and $\piz$s, which originates from the uncertainty
on the $\nn$ reference and $N_{coll}$.

To address suggestions that the yield of high $\pt$ hadrons in
$\Au$ collisions may be proportional to $\npart$ instead of
$\ncoll$~\cite{dima,mueller}, we have investigated a different
ratio,
\begin{equation}
 \raap = 2\ancoll/\anpart\times\raa \qquad .
\end{equation}
$\raap$ for $\pt>4.5$ GeV/$c$ is shown in the lower panel of
Figure~\ref{fig:Raainte}, together with solid (or dashed) bands
representing the allowed range if the data follow binary collision
(or participant) scaling. As discussed above, for peripheral
collisions, $\raap$ follows more closely the binary collision
scaling. Above 50 participants, $\raap$ varies by only $\pm$20\%.
However, it peaks at $\anpart$ = 100 and decreases monotonically
towards more central collisions~\footnote{In the $\pt$ range from
3--4 GeV/$c$, $\raap$ for charged hadrons is approximately
constant, which is consistent with earlier measurements at $\snn$
= 130 GeV~\cite{ppg013} and $\snn$ = 200 GeV~\cite{phobosch}. To
interpret this constancy as participant scaling is misleading,
since pion and proton yields change differently with centrality in
this $\pt$ region, and $\raap$ accidentally appears constant for
inclusive charged hadron. The data above 4.5 GeV/$c$ shown in
Figure~\ref{fig:Raainte} are free of this effect.} .

The decrease of $\raap$ could be a natural consequence of energy
loss of hard scattered partons in the medium~\cite{mueller}. If
the energy loss is large, hard scattered partons may only escape
near the surface of the reaction volume. In a cylindrical
collision geometry, for which the number of collisions from the
surface is proportional to $\npart$, binary collision scaling is
reduced to an approximate participant scaling. Detailed
calculations show that in this case, $\raap$ slightly decreases
with $\npart$ depending in details on how the energy loss is
modelled~\cite{mueller}. This interpretation is also consistent
with our previous conclusion that, above 4.5~GeV/$c$, hadron
production is dominated by hard-scattering although the yield does
not scale with the number of binary collisions. Gluon saturation
scenarios~\cite{dima} also suggest approximate participant
scaling, with a 30\% increase in $\raa$ over the $\pt$ range
4.5--9 GeV/$c$ in central collisions. This increase can not be
excluded by the data.
\subsection{Energy dependence and $x_T$ scaling} \label{sec:SxT}

The inclusive charged hadron and $\piz$ $\pt$ spectra and $h/\piz$
ratios suggest that fragmentation of hard scattered partons is the
dominant production mechanism of high $\pt$ hadrons not only in
$\pp$ but also in $\Au$ collisions. For $\pp$ collisions this fact
was demonstrated on general principles well before the advent of
QCD by the method of ``$x_T$-scaling''. This method does not
depend on whether the initial projectiles are protons or $Au$
ions, so it should be directly applicable to $\Au$ collisions.
Since our data show a suppression of high-$p_T$ particles in
central $\Au$ collisions with respect to point-like scaling from
$\pp$ and peripheral $\Au$ collisions, it is important to
investigate whether the production dynamics of high-$p_T$
particles in central (and peripheral) $\Au$ collisions are the
same or different from those in $\pp$ collisions. We first review
the $x_T$-scaling method in $\pp$ collisions and then apply it to
the present $\Au$ data.

The idea of hard-scattering in $\nn$ collisions dates from the
first indication of point-like structure inside the proton, in
1968, found in deep inelastic electron-proton
scattering~\cite{DIS}, i.e. scattering with large values of
4-momentum transfer squared, $Q^2$, and energy loss, $\nu$. The
discovery that the Deep Inelastic Scattering (DIS) structure
function
\begin{equation}
F_2(Q^2, \nu)=F_2\left({Q^2 \over \nu}\right) \label{eq:F2scales}
\end{equation}
 ``scales'', or in other words, depends on the ratio
\begin{equation}
x=\frac{Q^2}{2M\nu} \label{eq:x_def}
\end{equation}
independent of $Q^2$ as suggested by Bjorken~\cite{Bj}, led to the
concept of a proton being composed of point-like ``partons''.
Since the partons of DIS are charged, and hence must scatter
electromagnetically from each other in $\pp$ collisions, a general
formula for the cross section of the single-particle inclusive
reaction
\begin{equation}
 p + p\rightarrow C +X
\label{eq:bbk1}
\end{equation}
was derived~\cite{BBK} using the principle of factorization of the
reaction into parton distribution functions for the protons,
fragmentation functions to particle $C$ for the scattered partons
and a short-distance parton-parton hard scattering cross section.

The invariant cross section for the single-particle inclusive
reaction (Eq.~\ref{eq:bbk1}), where particle $C$ has transverse
momentum $p_T$ near mid-rapidity, was given by the general scaling
form~\cite{CIM}:
\begin{equation}
E \frac{d^3\sigma}{dp^3}=\frac{1}{p_T^{n}} F\left({2 p_T \over
\sqrt{s}}\right) \quad \mbox{where}\quad x_T=2p_T/\sqrt{s} \qquad
. \label{eq:bbg}
\end{equation}
The cross section has 2 factors: a function $F$ which depends only
on the ratio of momenta, and a dimensioned factor, ${p_T^{-n}}$,
where $n$ depends on the quantum exchanged in the hard-scattering.
For QED or Vector Gluon exchange~\cite{BBK}, $n=4$. For the case
of quark-meson scattering by the exchange of a quark~\cite{CIM},
$n$=8. The discovery of high $p_T$ pions in $\pp$ scattering at
the CERN-ISR, in 1972~\cite{CCR,SS,BS}, at a rate much larger than
predicted by electromagnetic scattering, but with the scaling form
of Eq.~\ref{eq:bbg},  proved that the partons of DIS strongly
interact with each other.

Inclusion of QCD~\cite{CGKS} into the scaling form led to the
$x_T$-scaling law (Eq.~\ref{eq:bbg}),
\begin{equation}
E \frac{d^3\sigma}{dp^3}={1\over {\sqrt{s}^{{\,n(x_T,\sqrt{s})}} }
} \: G({x_T}) \qquad, \label{eq:nxt}
\end{equation}
where the ``$x_T$-scaling power'' $n(x_T,\sqrt{s})$ should equal 4
in lowest order (LO) calculations, analogous to the $1/q^4$ form
of Rutherford Scattering in QED. The structure and fragmentation
functions, which scale as the ratios of momenta are all in the
$G(x_T)$ term. Due to higher order effects such as the running of
the coupling constant, $\alpha_s(Q^2)$, the evolution of the
structure and fragmentation functions, and the initial state
$k_T$, measured values of ${\,n(x_T,\sqrt{s})}$ in $\pp$
collisions are in the range from 5 to 8.

\begin{figure*}[ht]
\includegraphics[width=0.40\linewidth]{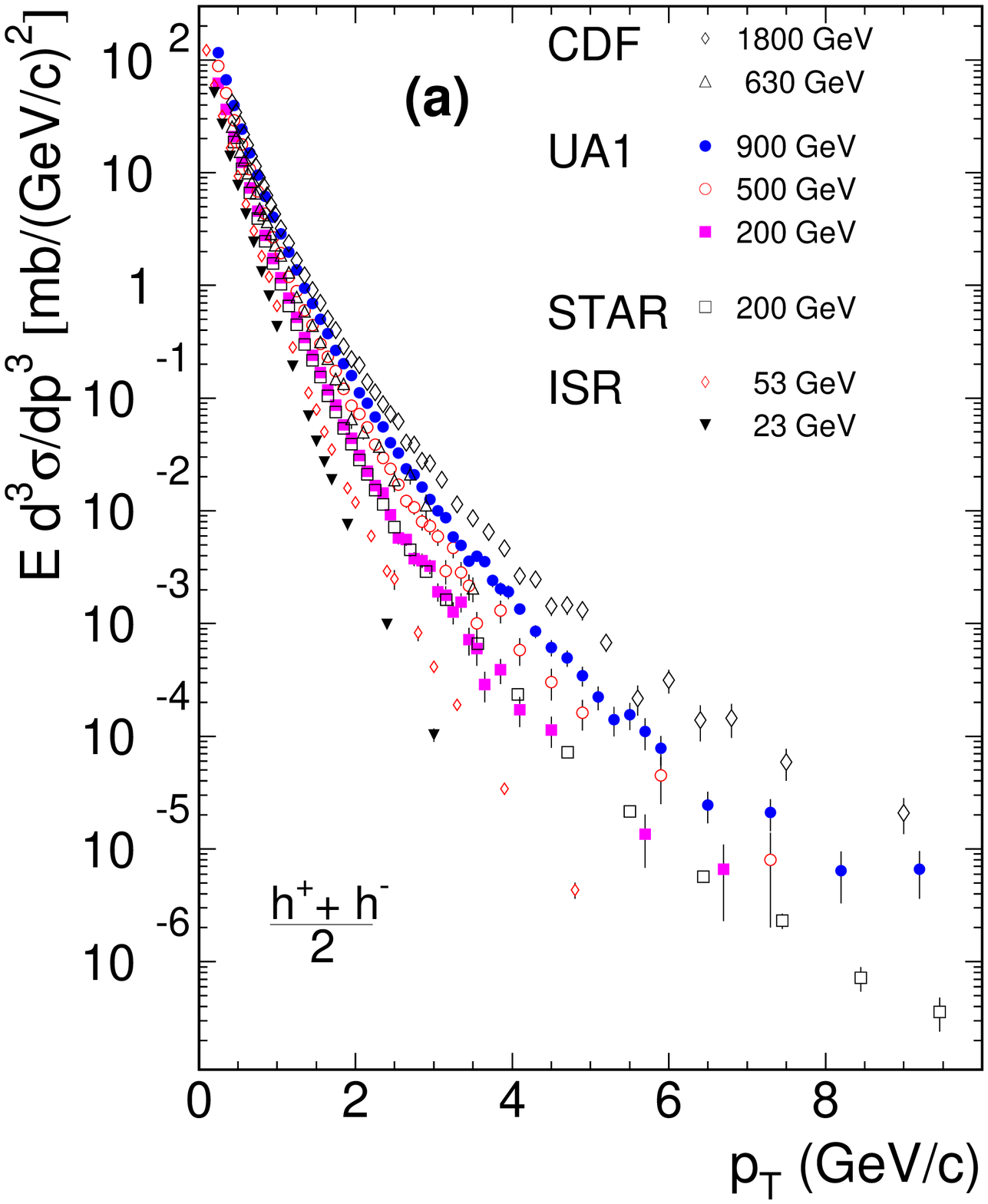}
\includegraphics[width=0.40\linewidth]{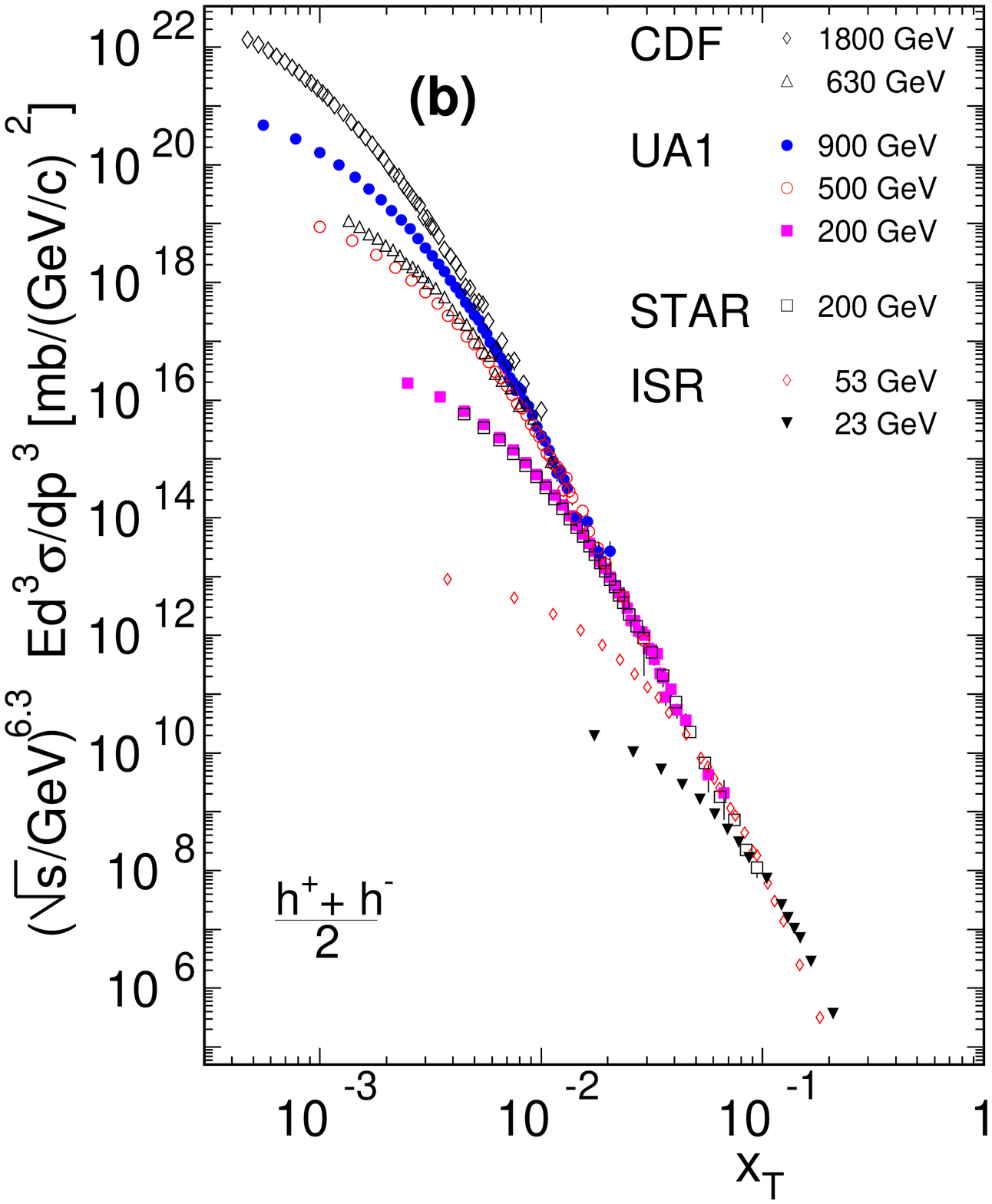}
    \caption{
    (Color online) (a) Transverse momentum dependence of the
    invariant cross section at seven center of mass energies
    from different experiments~\cite{ISR,CDF,UA1,STAR}. (b) The
    same data multiplied by $\sqrt{s}^{\,6.3}$, plotted as a
    function of $x_T=2 p_{T}/\sqrt{s}$. \label{fig:pp-h}}
\end{figure*}

\begin{figure*}[ht]
\includegraphics[width=0.40\linewidth]{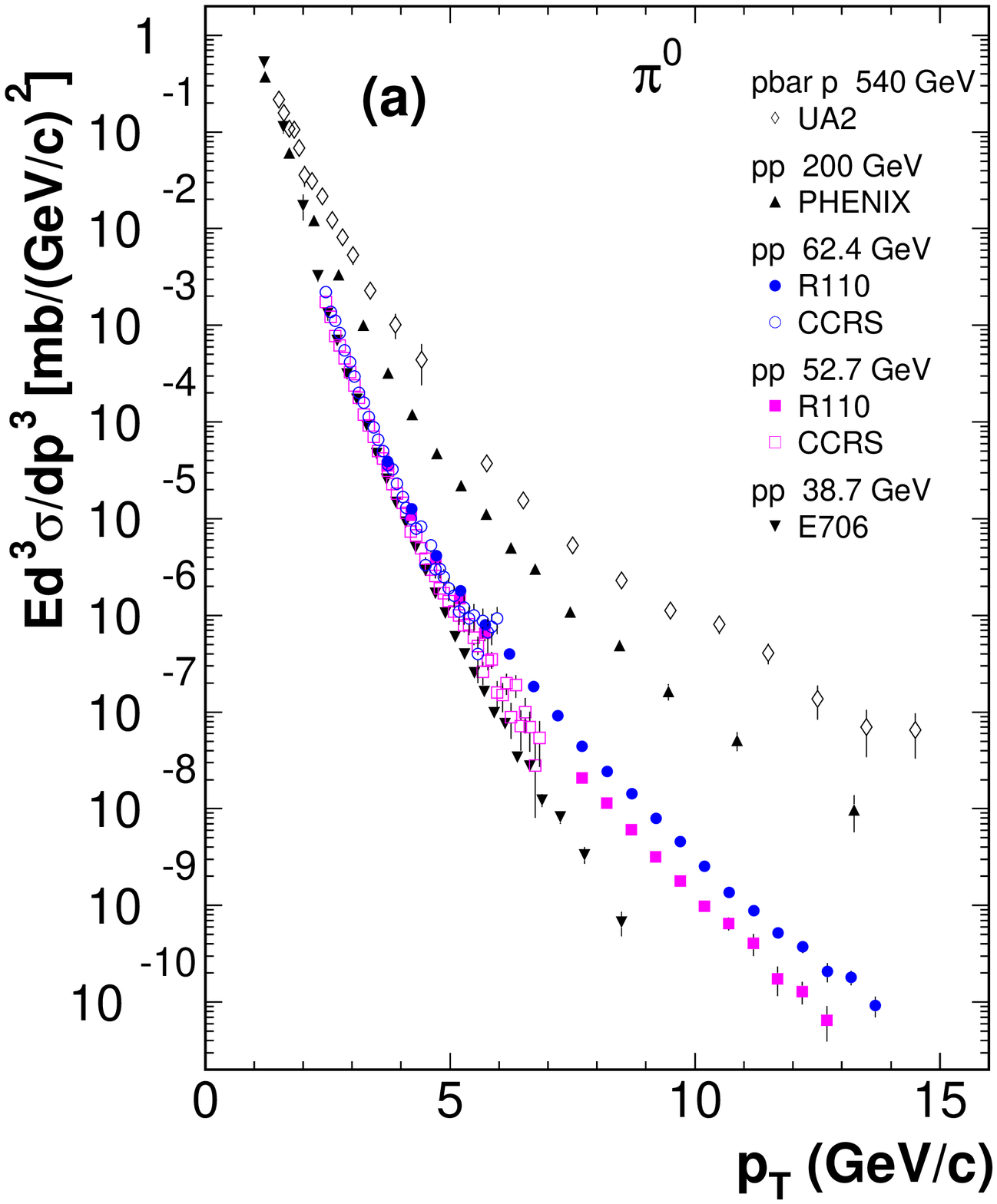}
\includegraphics[width=0.40\linewidth]{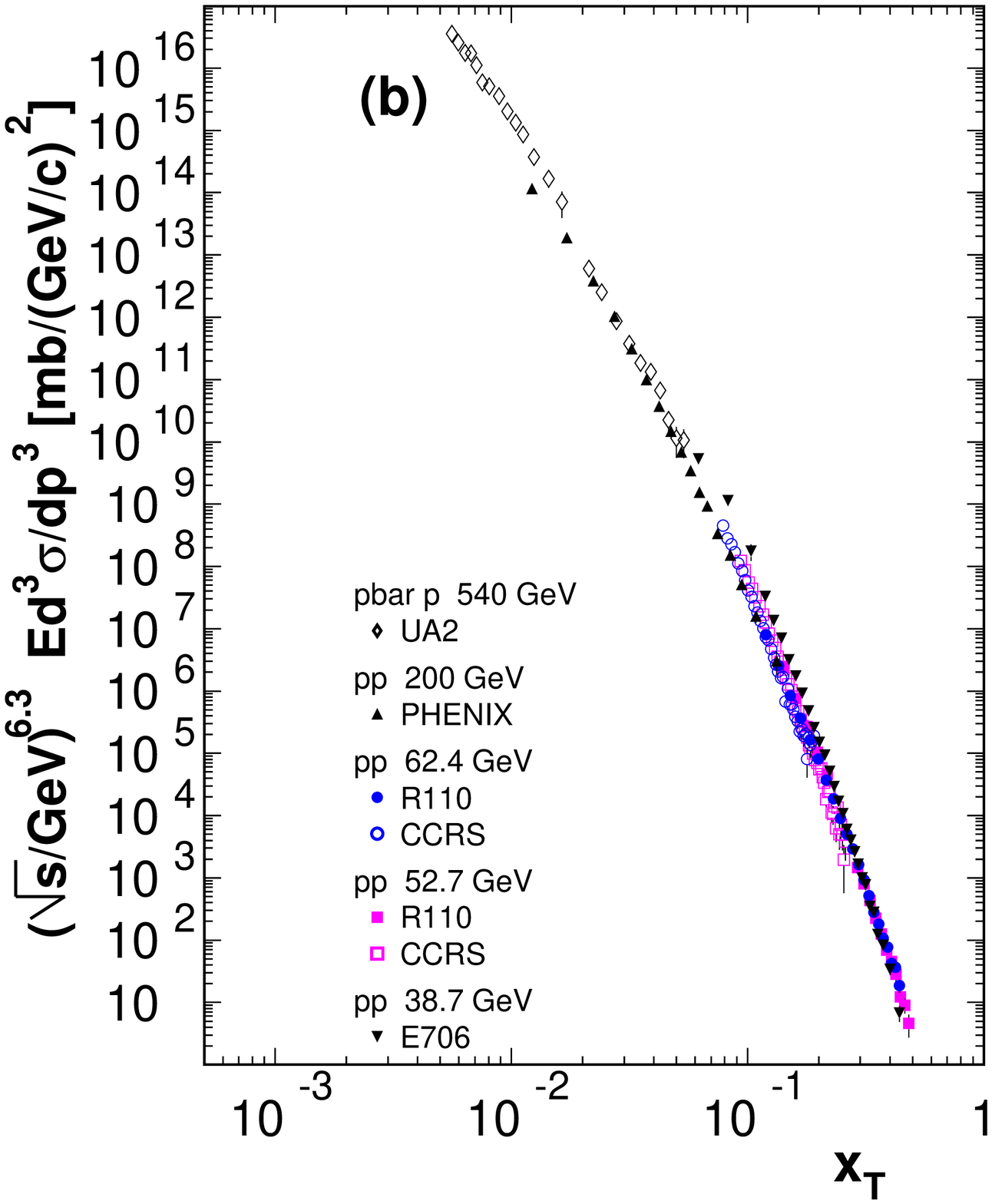}
\caption {
    (Color online) (a) Transverse momentum dependence of the invariant
     cross section for $\pi^0$ at five center-of-mass energies
     from different experiments~\cite{CCOR,CCRS,UA2,E706,ppg024}.
     (b) The same data multiplied by $\sqrt{s}^{\,6.3}$, plotted vs
     $x_T=2 p_{T}/\sqrt{s}$. \label{fig:pp-pi0} }
\end{figure*}

The compilation of single particle inclusive transverse momentum
spectra at mid-rapidity from $\pp$ and $p+\bar{p}$ collisions at
c.m. energy $\sqrt{s}$ from 23 to 1800 GeV~\cite{ISR,CDF,UA1,STAR}
is shown in Figure~\ref{fig:pp-h}a for $(h^{+} + h^{-})/2$, and in
Figure~\ref{fig:pp-pi0}a for
$\pi^0$~\cite{CCOR,CCRS,UA2,E706,ppg024}. The spectra exhibit a
characteristic shape: an exponential part at low $p_T\leq 1$
GeV/$c$ which depends very little on $\sqrt{s}$ (soft physics),
and a power-law tail for $p_T \geq 2$ GeV/$c$ which depends very
strongly on $\sqrt{s}$ (hard physics). The high $\pt$ part of the
spectra shows a characteristic scaling behavior indicative of
fragmentation of jets produced by hard-scattering of the quark and
gluon constituents of the proton as described by
QCD~\cite{Dilella,Darriulat,owens2}.

The $\xt$-scaling of the single particle inclusive data is nicely
illustrated by a plot of
\begin{equation}
 \sqrt{s}^{{\,n(x_T,\sqrt{s})}} \times E \frac{d^3\sigma}{dp^3}=G(x_T) \qquad ,
\label{eq:xTscaling}
\end{equation}
as a function of $x_T$, with ${{\,n(x_T,\sqrt{s})}}=6.3$. The
$(h^+ + h^-)/2$ data (Figure~\ref{fig:pp-h}b) show an asymptotic
power law with increasing $\xt$. Data at a given $\sqrt{s}$ fall
below the asymptote at successively lower values of $\xt$ with
increasing $\sqrt{s}$, corresponding to the transition region from
hard to soft physics in the $\pt$ range of 1--2 GeV/$c$. The
$\pi^0$ data (Figure~\ref{fig:pp-pi0}b) show a similar
$\xt$-scaling but without the deviation at low $\xt$, since all
available data are for $\pt$ larger than 1--2 GeV/$c$. For larger
$x_T\geq 0.3$, a value of $n=5.1$~\cite{CCOR,ABCS} improves the
scaling for the 3 lower c.m. energies, $\sqrt{s}=38.7$, 52.7 and
62.4 GeV. It will be a challenge at RHIC to obtain data in this
$x_T$ range to see whether the value of $n\sim5$ is the asymptotic
limit for inclusive single particle production or whether $n$
reaches the (LO) QCD value of 4. $x_T$-scaling has also been
studied in jet production at $\sqrt{s}=630$ and 1800
GeV~\cite{Huston},  where $n=4.45$ is observed in the jet $x_T$
range 0.15--0.3.

In $\Au$ collisions, $x_T$-scaling should work just as well as in
$\pp$ collisions and should yield the same value of
$n(x_T,\sqrt{s})$ if the high $p_T$ particles are the result of
hard-scattering according to QCD. This is because the structure
and fragmentation functions in $\Au$ collisions should scale, so
that Eq.~\ref{eq:nxt} applies, albeit with a different $G(x_T)$.
Thus, if the suppression of high-$p_T$ particles with respect to
point-like scaling from $\pp$ collisions is due to shadowing of
the structure functions~\cite{EKS98} or gluon
saturation~\cite{dima}, which are basically scaling
effects~\footnote{There is a slight non scaling effect of the
structure functions~\cite{EKS98} since for fixed $x_T$, $Q^2$
changes by a factor of 2.4 between the two $\snn$.}, rather than
due to a final state interaction with the dense medium, which may
not scale, the cross sections (Eq.~\ref{eq:nxt}) at a given $x_T$
(and centrality) should all exhibit the same suppression. The
initial state shadowing may cause $G({x_T})$ to change with
centrality, but $n(x_T,\sqrt{s})$ should remain constant. In the
case of the interaction with the dense medium, $x_T$-scaling may
or may not hold, depending on the details of the energy loss, for
instance, whether or not the energy loss of the hard-scattered
parton scales with its energy. It is also conceivable that the
high $p_T$ particles observed in $\Au$ collisions at RHIC have
nothing to do with QCD
hard-scattering~\cite{coalence,junction,gallmeister}. In this
case, striking differences from Eq.~\ref{eq:nxt} and the
systematics observed in $\pp$ collisions should be expected.

\begin{figure*}[ht]
\includegraphics[width=0.75\linewidth]{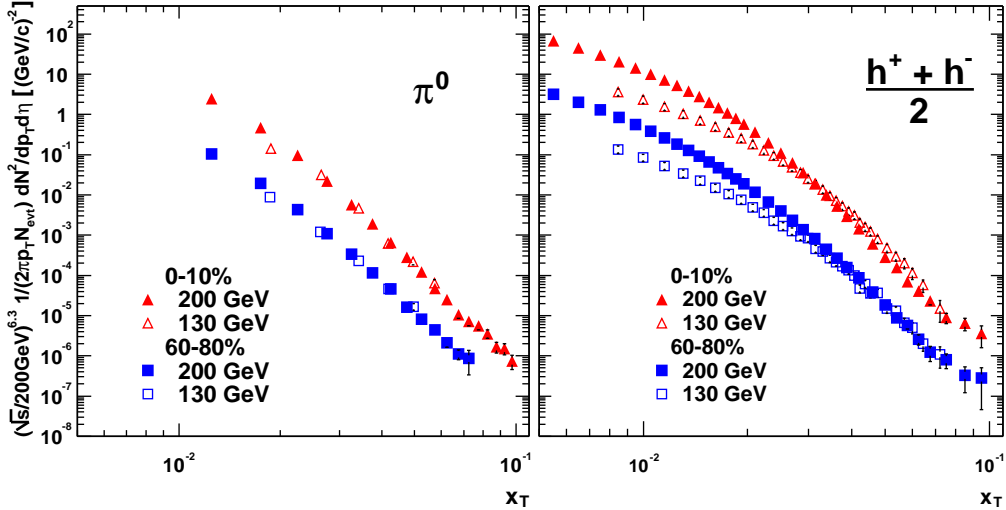}
     \caption{\label{fig:xT1}
    (Color online) $\xt$ scaled spectra for central collisions and peripheral
    collisions at $\snn$ = 130 and 200 GeV. The left figure shows the
    $\piz$ $\xt$ spectra, and the right figure shows the $(h^+ + h^-)/2$
    $\xt$ spectra. The central (0-10\%) $\xt$ spectra are represented by
    triangular symbols, and the peripheral (60-80\%) $\xt$ spectra are
    represented by square symbols. The open symbols represent
    $\xt$ spectra from $\snn$ = 130 GeV scaled by a factor of
    $(130/200)^{6.3}$. The solid symbols represent $\xt$ spectra
    from $\snn$ = 200 GeV. The error bars are statistical only.
   }
\end{figure*}

To test $\xt$-scaling in $\Au$ collisions, we plot the quantities
defined by Eq.~\ref{eq:xTscaling} in Figure~\ref{fig:xT1} for
charged hadron and $\piz$ data from $\snn$ = 130 GeV and 200 GeV
for central (0-10\%) and peripheral (60-80\%) collisions. For the
power $n$, we use the same value ${{\,n(x_T,\sqrt{s})}}=6.3$ that
was used for the $\pp$ data shown in Figure~\ref{fig:pp-h}b and
Figure~\ref{fig:pp-pi0}b. The data are consistent with
$\xt$-scaling over the range $0.03\leq x_T\leq 0.06$ for $\pi^0$
and $0.04\leq x_T\leq 0.075$ for $(h^{+} + h^{-})/2$.

\begin{figure*}[ht]
\includegraphics[width=0.75\linewidth]{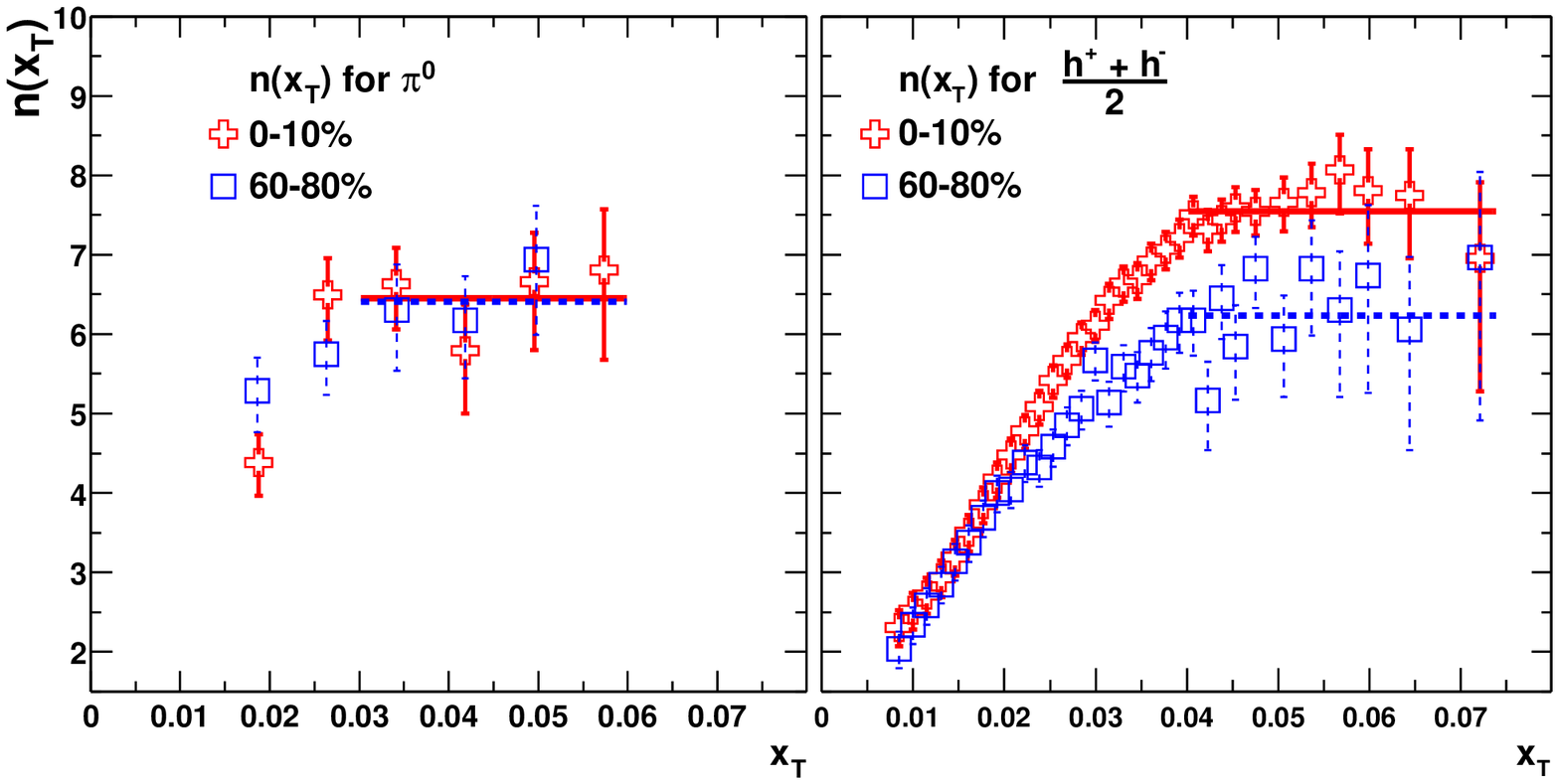}
   \caption{\label{fig:xT2}
   (Color online) The $\xt$ scaling power $n$ (according to Eq.~\ref{eq:n}) plotted as function of $\xt$ calculated
   for $\pi^0$ (top-left) and $(h^+ + h^-)/2$ (top-right) in central (0-10\%)
   and peripheral (60-80\%) collisions. The solid (and dashed) lines indicate
   a constant fit along with the fitting ranges to the central (and peripheral) $n(\xt)$ functions.
   The error bars at each data
   point include statistical and point-to-point systematic errors
   from $\snn$ = 130 and 200 GeV. The scale errors on $\xt$ spectra are 20.7\% (15.9\%) for
   $\piz$ $\xt$ spectra ratio in central (peripheral) collisions, and 18.6\% (15.7\%) for $(h^+ + h^-)/2$
   $\xt$ spectra ratio in central (peripheral) collisions. These type of errors propagate into
   the systematic errors on $\xt$ scaling power $n$ listed in Table~\ref{tab:xt}.
   }
\end{figure*}

According to Eq.~\ref{eq:nxt}, the ratio of inclusive cross
sections at fixed $\xt$ equals $(200/130)^{n}$. Thus, the power
$n(\xt,\sqrt{s})$ is related directly to the logarithm of the
ratio of invariant hadron yield at fixed $\xt$ as:
\begin{equation}
n(\xt) = \frac{log(yield(\xt,130 GeV)/yield(\xt,200
GeV))}{log(200/130)}.\label{eq:n}
\end{equation}
The power $n'$s for both neutral pions and charged hadrons for
central and peripheral collisions are shown in
Figure~\ref{fig:xT2}. While the $\piz$ data in central and
peripheral collisions and charged hadron data in peripheral
collisions seem to favor a similar power $n$, the charged hadron
data from central collisions require a larger value of $n$.

For a more quantitative analysis, the $\Au$ data for a given
centrality and hadron selection are fitted simultaneously for
$\snn$ = 130 and 200 GeV to the form,
\begin{equation}
   \left(\frac{A}{\sqrt{s}}\right)^n (x_T)^{-m} \qquad ,
\label{eq:10}
\end{equation}
where we have approximated Eq.~\ref{eq:nxt} by using a constant
power $n(x_T,\sqrt{s})$ and a power-law, $x_T^{-m}$, for $G(x_T)$
over a limited range in $\xt$. The fit results and errors are
quoted in Table~\ref{tab:xt}. The corresponding ratios of yields
are presented by lines in Figure~\ref{fig:xT2}, where the fit
ranges ($0.03\leq x_T\leq 0.06$ for $\piz$s and $0.04\leq x_T\leq
0.074$ for charged hadrons) are indicated by the length of the
line.

For peripheral collisions the fitted values for the power are
$n=6.33\pm0.54$ and  $n=6.12\pm0.49$, for $\piz$ and charged
hadrons respectively, which are in quantitative agreement with the
expectation from $\pp$ collisions. Approximate $\xt$-scaling in
peripheral $\Au$ collisions with the same power as observed in
$\pp$ collisions indicates that hard-scattering is the dominating
production mechanism for high $\pt$ particles. In central
collisions, neutral pions also exhibit $\xt$-scaling with a
similar power, $n=6.41\pm0.55$. Thus, it seems that high-$\pt$
$\piz$ production is consistent with hard-scattering, with scaling
structure and fragmentation functions, for all centralities.

For charged hadrons, the power found for central collisions is
$n=7.53\pm0.44$. Most of the systematic errors are common and
cancel between central and peripheral collisions, thus the
difference of the two powers found for charged hadrons, $\Delta n
= n_{cent}-n_{periph}= 1.41\pm0.43$ compared with that for neutral
pion $\Delta n = 0.09\pm0.47$, is significant.

This difference is consistent with the large proton and
anti-proton enhancement in central $\Au$ collisions for
intermediate $p_T$ seen at $\snn = 130$ and 200 GeV, which appears
to violate $\xt$-scaling. The $x_T$ range $0.04\leq x_T\leq 0.074$
corresponds to $4 < \pt < 7.4$ GeV/$c$ at $\sqrt{s_{NN}} = 200$
GeV, but it corresponds to $2.6 < \pt < 4.8$ GeV/$c$ at $\snn$ =
130 GeV. If protons are enhanced at $2 < \pt < 4.5$ GeV/$c$ in
central collisions at both $\snn$ = 130 GeV and 200 GeV, then
$n_{cent}$ will be larger than $n_{periph}$ in the measured $\xt$
range. Since $\snn$ = 200 GeV data indicate that the proton
enhancement is limited to the medium $\pt$ range, based on the
equality of $\raa$ for charged hadrons and $\pi^0$ at $p_T> 4.5$
GeV/$c$ (Figure~\ref{fig:Raa}), this difference should go away at
larger $\xt$ .

\begin{table}
\begin{center}
\caption{\label{tab:xt} Results of the simultaneous fit to $\snn$
= 130 and 200 GeV data using Eq.~\ref{eq:10}. The fit ranges are
$0.03\leq x_T\leq 0.06$ for $\piz$ and $0.04\leq x_T\leq 0.074$
for charged hadron. Only statistical and point-to-point systematic
errors on the data points are included in the fit, which gives the
statistical error on $n$. The normalization errors and other $\pt$
correlated systematic errors are not included in the fit but are
directly translated into a systematic error on $n$.}
  \begin{ruledtabular}  \begin{tabular}{rrr}
\multicolumn{3}{c}{Fitting results for $\pi^0$ over $0.03<\xt<0.06$}\\ \hline
parameters & 0-10\% centrality bin & 60-80\% centrality bin\\ \hline
$A$ & $0.973\pm0.232$     & $0.843\pm0.3$\\
$m$ & $8.48\pm0.17$       & $7.78\pm0.22$\\
$n$ & $6.41\pm0.25(stat)$ & $6.33\pm0.39(stat)$\\
    & $\pm0.49(sys)$      & $\pm0.37(sys)$\\ \hline
\multicolumn{3}{c}{Fitting results for $h^{+} + h^{-}$ over $0.04<\xt<0.074$}\\ \hline
$A$ & $2.30\pm0.44$       & $0.62\pm0.27$\\
$m$ & $8.74\pm0.28$       & $8.40\pm0.43$\\
$n$ & $7.53\pm0.18(stat)$ & $6.12\pm0.33(stat)$\\
    & $\pm0.40(sys)$      & $\pm0.36(sys)$\\
\end{tabular}  \end{ruledtabular}
\end{center}
\end{table}

\section{SUMMARY}
\label{sec:conclusion}

We have presented a systematic study of the $\pt$ and centrality
dependence of charged hadron production at $|\eta|<$0.18 at $\snn$
= 200 GeV. The yields per nucleon-nucleon collision in central
collisions are significantly suppressed compared to peripheral and
nucleon-nucleon collisions. The suppression is approximately
independent of $\pt$ above 4.5 GeV/$c$ for all centrality classes,
suggesting a similar spectral shape between $\Au$ and $\pp$
collisions. At $\pt>4.5$, charged hadron suppression is the same
as for neutral pions; the ratio $h/\piz$ is $\sim1.6$ for all
centralities, similar to the $h/\pi$ value measured in $\pp$ and
$e^+e^-$ collisions. The similar spectral shape and particle
composition at high $\pt$ are consistent with jet fragmentation as
the dominating mechanism of particle production in $\Au$
collisions for $\pt>$ 4--5 GeV/$c$. For both charged hadrons and
neutral pions, the suppression sets in gradually from peripheral
to central collisions, consistent with the expectation of partonic
energy loss and surface emission of high $\pt$ hadrons.
$\xt$-scaled hadron yields are compared between $\snn$ = 130 GeV
and $\snn$ = 200 GeV $\Au$ collisions. We find that the $\xt$
scaling power $n$ calculated for neutral pions in central and
peripheral collisions and charged hadron in peripheral collisions
is $6.3\pm0.6$, similar to $\pp$ collisions. This again points
towards similar production dynamics, i.e. hard-scattering
processes as described by QCD. However, $n$ is $7.5\pm0.5$ for
charged hadrons in central collisions, indicating a strong
non-scaling modification of particle composition of charged hadron
spectra from that of $\pp$ at intermediate $\pt$, 2--4.5 GeV/$c$.
This is consistent with the large $h/\piz$ ratios observed over
the same $\pt$ range in central collisions.

\begin{acknowledgments}


We thank the staff of the Collider-Accelerator and Physics
Departments at Brookhaven National Laboratory and the staff
of the other PHENIX participating institutions for their
vital contributions.  We acknowledge support from the
Department of Energy, Office of Science, Nuclear Physics
Division, the National Science Foundation, Abilene Christian
University Research Council, Research Foundation of SUNY,
and Dean of the College of Arts and Sciences, Vanderbilt
University (U.S.A), Ministry of Education, Culture, Sports,
Science, and Technology and the Japan Society for the
Promotion of Science (Japan), Conselho Nacional de
Desenvolvimento Cient\'{\i}fico e Tecnol{\'o}gico and
Funda\c c{\~a}o de Amparo {\`a} Pesquisa do Estado de
S{\~a}o Paulo (Brazil), Natural Science Foundation of China
(People's Republic of China), Centre National de la
Recherche Scientifique, Commissariat {\`a} l'{\'E}nergie
Atomique, Institut National de Physique Nucl{\'e}aire et
de Physique des Particules, and Institut National
de Physique Nucl{\'e}aire et de Physique des Particules,
(France), Bundesministerium fuer Bildung und
Forschung, Deutscher Akademischer Austausch Dienst, and
Alexander von Humboldt Stiftung (Germany), Hungarian
National Science Fund, OTKA (Hungary), Department of Atomic
Energy and Department of Science and Technology (India),
Israel Science Foundation (Israel), Korea Research
Foundation and Center for High Energy Physics (Korea),
Russian Ministry of Industry, Science and Tekhnologies,
Russian Academy of Science, Russian Ministry of Atomic Energy
(Russia), VR and the Wallenberg Foundation (Sweden), the U.S.
Civilian Research and Development Foundation for the Independent
States of the Former Soviet Union, the US-Hungarian NSF-OTKA-MTA,
the US-Israel Binational Science Foundation, and the 5th
European Union TMR Marie-Curie Programme.

\end{acknowledgments}


\end{document}